%% file: NLSMstring.tex
\input harvmacM.tex
\input epsf.sty
\input amssym.tex 
\input color

\def\dd{{\rm d}}
\def\frac#1#2{{#1\over #2}}

\title Abelian $Z$-theory:

\title NLSM amplitudes and $\alpha'$-corrections from the open string

\author
John Joseph M. Carrasco$^{a,b}$\email{1}{john-joseph.carrasco@cea.fr},
Carlos R. Mafra$^{c,b}$\email{2}{c.r.mafra@soton.ac.uk} and
Oliver Schlotterer$^d$\email{3}{olivers@aei.mpg.de}

\address
$^a$ Institut de Physique Th\'eorique,
CEA--Saclay, F--91191 Gif-sur-Yvette cedex, France
\medskip
$^b$Institute for Advanced Study, School of Natural Sciences,
Einstein Drive, Princeton, NJ 08540, USA
\medskip
$^c$ STAG Research Centre and Mathematical Sciences,
University of Southampton, Highfield, Southampton SO17 1BJ, UK
\medskip
$^d$Max--Planck--Institut f\"ur Gravitationsphysik,
Albert--Einstein--Institut,
Am M\"uhlenberg 1, 14476 Potsdam, Germany

\abstract
In this paper we derive the tree-level S-matrix of the effective theory of
Goldstone bosons known as the non-linear sigma model (NLSM) from 
string theory. This novel connection  relies on a recent realization of tree-level 
open-superstring S-matrix predictions as a 
{\it double copy} of super-Yang--Mills theory
with $Z$-theory --- the collection of putative scalar effective field theories 
encoding all the $\ap$-expansion of the open superstring.  
 Here we
identify the color-ordered amplitudes of the NLSM  as the low-energy
limit of abelian $Z$-theory.  This realization also provides
natural higher-derivative corrections to the NLSM amplitudes arising from 
higher powers of $\ap$ in the abelian $Z$-theory amplitudes, and through double copy 
also to Born--Infeld and Volkov--Akulov theories. The amplitude relations due to Kleiss--Kuijf as well as
Bern, Johansson and one of the current authors obeyed by $Z$-theory amplitudes thereby 
apply to all $\ap$-corrections of the NLSM. As such we naturally obtain 
a cubic-graph parameterization for the abelian $Z$-theory predictions 
whose kinematic numerators obey the duality between color and kinematics 
to all orders in $\ap$.



\input refs

\listtoc
\writetoc
\filbreak

\newsec Introduction

It is well known that string theory provides a powerful and unified framework to
study the sea of field theories that arise in the limit when the size
of the strings approaches zero; some of the most celebrated examples
being the maximally supersymmetric super-Yang--Mills and supergravity theories~\GreenFT.  

Since the gluon and graviton belong to the massless excitations of the string,
their scattering amplitudes naturally emerge from low-energy limits of the
string-theory S-matrix. By the same token, one might suspect that scattering
amplitudes of field theories absent in the naive\foot{Given the string-theory realizations 
of principle chiral models through toroidal compactifications along with worldsheet 
boundary condensates \GreenGA, the term ``naive string spectrum'' refers to string 
theories in $D$-dimensional uncompactified Minkowski spacetime.}
string spectrum may be difficult
to study within string theory.  As we will see, such an expectation is surpassed by
a long-hidden double-copy structure, secretly and deftly encoded in open-string
theory---a structure which applies universally to a broad set of point-like 
quantum field theories~\Ztheory.  Perturbative predictions in double-copy quantum
field theories~\refs{\KLT, \BCJ, \CHY}
can be completely fixed by knowing the
predictions of two possibly distinct input theories\foot{For
example, scattering in the ${\cal N}=5$ supergravity theory is completely determined
by a double copy consisting of
color-stripped amplitudes of ${\cal N}=4$ and ${\cal N}=1$ super Yang--Mills theories.
Note that double copy holds at the integrand level~\BCJLoops\ for multiloop
amplitudes.}.

In recent work by Broedel, Stieberger, and one of the current authors \Ztheory\
it was demonstrated that open-superstring amplitudes \refs{\nptTreeI,\nptTreeII}
can be understood as a double copy of color-stripped Yang--Mills amplitudes and
certain $Z$-functions which behave like scalar partial amplitudes.  These
$Z$-functions are iterated integrals over the boundary of a disk worldsheet and
naturally incorporate two notions of ordering.  One ordering, $Q=(q_1,q_2, \ldots,q_n)$, 
refers to the integrand,  
and one to the
integration domain $P=(p_1,p_2, \ldots,p_n)$, (see section~2.2)
\eqn\ZintdefIntro{
Z_{P}(q_1,q_2,\ldots,q_n) \equiv \ap^{n-3}\!\!\!\!
\int \limits_{D(P)} {\dz_1 \ \dz_2 \ \cdots  \ \dz_n \over {\rm vol}(SL(2,\Bbb R))}
 { \prod_{i<j}^n |z_{ij}|^{\ap s_{ij}}  \over z_{q_1 q_2} z_{q_2 q_3} \ldots z_{q_{n-1} q_n} z_{q_n q_1}}\,.
}
It was shown in ref.~\Ztheory\ that these
doubly-ordered functions obey Kleiss--Kujif (KK) \KKsym\ and
Bern--Carrasco--Johansson (BCJ) \BCJ\ field-theory amplitude relations along its
integrand ordering~$Q$, and the string-theory monodromy
relations~\refs{\BjerrumBohrRD, \StiebergerHQ} along the integration domain~$P$.
After dressing the string-monodromy ordering $P$ with the
appropriate sum over string Chan--Paton factors, we are left with the
$Q$-ordered functions that simply obey field-theory relations. Could
these functions be {\it color}-ordered scattering amplitudes for
some effective field theory?  We conjecture the answer is yes. In
this manuscript we provide non-trivial evidence in the abelianized
low-energy limit, discovering the $D$-dimensional color-ordered
amplitudes of the non-linear sigma model (NLSM), shown to satisfy
the relevant color (flavor)-ordered field-theory scattering
amplitude relations in ref.~\ChangZZA.

The KK and BCJ field-theory amplitude relations between ordered
scattering amplitudes are incredibly constraining and very
special---key to consistency properties of the theories they
describe.  The fact that the $Z$-functions \ZintdefIntro\ obey KK and BCJ
relations along an ordering suggests the possible interpretation
that these functions might themselves be scattering amplitudes for
some conjectural and as yet to be identified doubly-colored theory,
which we will refer to as $Z$-theory.  As such we refer to the
$Z$-functions as $Z$-amplitudes. This relies on a natural
conjecture about factorization properties of these functions supported
by the double-copy perspective as well as an open-string
worldsheet model which we will describe in more detail in
Appendix~A.

In the time-honored tradition of periodically 
reflecting upon the venerable query, ``What is string theory?'', we find ourselves struck
by the ubiquity of double-copy constructions -- not only in unifying open- and
closed-string predictions a la Kawai, Lewellen and Tye (KLT) \KLT, but also in constraining the effective-field-theory (EFT)
modifications to super-Yang--Mills resulting in open-superstring tree-level predictions.
We suspect a strategic path forward may arise from 
the more modest question driving this current manuscript, ``What is $Z$-theory?''.

A critical first clue was given in \Ztheory, where it was demonstrated that
the $\alpha'\to0$ limit of doubly-ordered $Z$-functions lands on the inverse of the field-theory
KLT matrix.
This same limit was later recognized to correspond to the (doubly-partial) tree-level 
amplitudes of a scalar bi-adjoint theory \DPellis. Therefore, even though
the string spectrum does not include 
bi-adjoint scalars, string tree-level amplitudes, through $Z$-amplitudes, contain all their tree-level predictions.

To be clear, we are {\it not} uplifting a known field theory's
predictions to string-theory amplitudes using some procedure.  
Rather these conjectural $Z$-amplitudes are double-copy factors in
open-superstring scattering amplitudes which contain all-orders in
$\ap$ and obey field-theory scattering amplitude relations.  They
are entirely well-defined functions sitting within the open-string
predictions. The scalar amplitudes of the conjectural $Z$-theory
retain the fingerprints of string-theory relevance through their
second Chan--Paton dressed monodromy-ordering. If our conjecture
holds, the (low-energy) $\ap$-expansion of these
objects should be identifiable as scattering amplitudes in some
identifiable field theories, with each successive order in $\ap$
representing higher-derivative corrections. 

Indeed, our primary result in this manuscript is the identification of such a
theory: the tree-level S-matrix of Goldstone bosons described by the $D$-dimensional
NLSM can be obtained from open-superstring amplitudes. Denoting the
Lie-algebra valued Goldstone-boson field by $\varphi$, the NLSM
Lagrangian in the Cayley parametrization is given by
\eqn\Lag{
{
{\cal L}_{{\rm NLSM}} = {1\over 2} {\rm Tr} \bigg\{
\partial_\mu \varphi \, {1\over 1-  \varphi^2} \, \partial^\mu \varphi  \, {1\over 1-   \varphi^2} 
\bigg\} 
}
}
with Lorentz indices $\mu=0,1,\ldots,D{-}1$. This theory's predictions emerge when the
Chan--Paton dressings of the $Z$-functions trivialize, i.e.\ are taken
to be abelian, and we consider the leading surviving order in $\ap$.
In other words, abelian $Z$-amplitudes arise from the tree-amplitudes
of abelian open-string states by replacing
Yang--Mills factors with gauge-theory color-factors. 
%
%
Specifically we find that NLSM amplitudes are given by\foot{In order to avoid cluttering of factors of two, we
have rescaled $\alpha'$ such that the standard open-string conventions are
recovered by setting $\alpha'\rightarrow 2\alpha'$ in the equations of this work.}
\eqn\preview{ A_{\rm NLSM}(1,2,\ldots,n) =
\lim_{\alpha' \rightarrow 0} (\ap)^{2-n}\!\!\!\!
\sum_{\sigma \in S_{n-1}} Z_{1\sigma(2,3,\ldots,n)}(1,2,\ldots,n)\,.
}
In contrast to the doubly-partial amplitudes of bi-adjoint scalars,
the NLSM amplitudes arise from nonzero orders $(\ap)^{n-2}$ singled
out by the leading low-energy contribution to $n$-point disk
integrals \ZintdefIntro\ in absence of an ordering\foot{For the fixed
field-theory color-ordered $Z$-theory amplitudes one dresses all
monodromy orderings with relevant Chan--Paton factors -- a sum over
$(n{-}1)!$ orders in the naive trace basis.  In the abelian case we
take the Chan--Paton factors trivial, effectively symmetrizing over
the integration order of the $Z$-functions, yielding color-ordered
abelian $Z$-theory amplitudes.} in the integration domain.
At four and six points, for instance, \preview\ yields
\eqnn\NLSMfour \eqnn\NLSMsix
$$\eqalignno{
A_{\rm NLSM}(1,2,3,4) &= \pi^2(s_{12}+s_{23})\,,
&\NLSMfour \cr
A_{\rm NLSM}(1,2,\ldots,6) &= \pi^4 \Bigl[ s_{12}
-\half { (s_{12}+s_{23}) (s_{45}+s_{56}) \over s_{123}}  + {\rm
cyclic}(1,2,3,4,5,6)\Bigr]\,.
&\NLSMsix
}$$
As will be explained below, the sum over string integrals in
\preview\ appears in
the $n$-point amplitude of massless open-string states
obtained in \refs{\nptTreeI,\nptTreeII} upon specialization to {\it abelian} Chan--Paton degress of freedom.  
Notice that abelianization refers only to the monodromy ordering, leaving the field-theory ordering
to be dressed with field-theory adjoint weights to represent color/flavor in the full dressed abelian
Z/NLSM amplitudes.

The surviving low-energy limit of these amplitudes are color-ordered
NLSM amplitudes. In addition, the subleading terms in the
$\ap$-expansion of \preview\ will be interpreted as stringy
higher-derivative corrections to the NLSM amplitudes. This
summarizes our primary result for which we report on explicit
calculations through nine points. We additionally provide a general
approach to generating associated local color-kinematic satisfying
numerators from these amplitudes.  Along the way we provide a nice
recursive form of the KLT matrix $S[\cdot | \cdot]_1$, inspired by
ref.~\DuTBC, and propose a strikingly simple form for
all-multiplicity $n=2k$ NLSM color-kinematic satisfying master
numerators
\eqnn\entireexpression
$$\eqalignno{
N_{1|\rho(23\ldots 2k-1)|2k} &= (-1)^{k} S[ \rho(23\ldots 2k{-}1) | \rho(23\ldots 2k{-}1) ]_1 \ . &\entireexpression
}$$
This manuscript is organized as follows: After a review of disk integrals in section 2,
we state and prove our main result \preview\ in section 3. Examples and the
systematics of higher-derivative corrections to the NLSM are elaborated in section
4, and the construction of explicit BCJ numerators for the $\alpha'$-corrected NLSM
can be found in section 5.

\newsec Review 
\par\seclab\sectwo

\subsec Double-copy construction
\par\seclab\secDCreview

\noindent Due to the seminal work~\KLT\ of KLT, it has
long been recognized that the tree-level predictions of open strings entirely
encode\foot{Given the all-multiplicity tree-level relations between gravity and
gauge theory, one might rightfully ask if classical general-relativity solutions
are encoded in classical gauge-theory solutions.  Answering this question is an
active area of investigation see e.g.\ refs.~\MonteiroCDA\
for explicit solution relationships, and refs.~\BorstenBP\
for considerations of classical symmetries and
duality-groups, as well as references therein.} the tree-level predictions of closed strings. 
The amplitudes admit a representation in the form of the sum over
products of color-stripped gauge (open-string) amplitudes.  This made a particular
impact in its low-energy limit in the study of field-theory gravitational
scattering amplitudes in the 90's where various closed-form representations were
identified \refs{\BerendsZP, \BernKLT}. 

Admittedly, many properties of the KLT double-copy construction were mysterious
(including the emergence of the necessary permutation symmetry of the gravity
amplitudes), and the original formulation of sums over shuffle-ordered products of
permutations could be dizzying, especially at higher multiplicities. Nevertheless,
this approach, as it would factor over the state-sum of unitarity cuts, proved
critical for gathering information about the spectacular ultraviolet behavior of the
maximally supersymmetric supergravity theory through
four-loops~\gravUVTwo.
In the process of such explorations, the field of scattering amplitudes acquired a new
set of insights relevant to the double-copy story when Bern, Johansson, and one of
the current authors (BCJ) observed~\BCJ\ that there was a very direct path to
gravity-theory predictions, where the double-copy construction could be made manifest
graph by graph.  This has particular value at the multiloop integrand
level~\BCJLoops, where integrand labeling ambiguities can create obstacles for
realizing generalized-gauge-invariant double-copy relationships, outside of certain
kinematic limits like unitarity cuts.

The BCJ double-copy approach relies critically on the realization~\BCJ\ that gauge-theory 
predictions (and their supersymmetric partners) admit a color-kinematic
duality satisfying representation (color-dual representations are ones where graph
by graph color-weights and kinematic weights obey the same generic algebraic
properties).
The existence of such color-dual
representations resulted in the discovery of new relations between color-ordered
amplitudes known now as the BCJ relations.  With such a dual
representation, color-factors could be consistently
replaced\foot{Established to some finite multiplicity at tree-level in \BCJ, but
later proven via BCFW and the KLT relations in \BernYG.} by kinematic weights,
recycling a small set of kinematic predictions
to describe a wide variety of theories~\BCJLoops.  Additionally, color-dual
kinematic weights could be solved for in terms of color-ordered amplitudes, thus
allowing for the generation of generalized KLT relations.

While a Lagrangian understanding of the organizing principle is only available in
the four-dimensional self-dual case~\SelfDualBCJ, many theories, including the NLSM
\chendu, in a variety of spacetime dimensions, admit the duality between color and
kinematics, and associated double-copy construction~\BargheerGV.
This new perspective on field-theory predictions has proven critical in
developing aspects of our understanding of non-planar scattering amplitudes over
the last decade, both formally as well as through practical reach in computation.
Jacobi relations drastically constrain the independent information
relevant to a given scattering calculation. For instance, the closely related
double-copy constructions of multiloop gravity amplitudes \BCJLoops\ has allowed
many explicit calculations that can probe the possible onset of ultraviolet
divergences in supergravity theories \BernUF.

String theory continues to provide key insights probing the color-kinematics
duality and its associated representation of the double-copy construction: from the
powerful proof \refs{\BjerrumBohrRD,\StiebergerHQ} of the $(n{-}3)!$-basis of
Yang--Mills tree amplitudes as the low-energy limit of the related string monodromy
relations\foot{See \loopMonodromy\ for a recent higher-loop generalization.},
to the elegant construction of explicit local tree-level numerators
\refs{\PSBCJ, \FTlimit}, to the construction of string-inspired BCJ
numerators~\MafraGJA\ at loop level.  The fact that the BCJ-duality also applies to
the NLSM \chendu, can now be appreciated either as a {\it consequence} of the BCJ
relations satisfied~\Ztheory\ by $Z$-theory as in \preview, or as a {\it requirement}
for the NLSM to be able to participate in $Z$-theory's construction of the open
superstring.  Following the recent result of Du and Fu \DuTBC\ who present an
elegant closed-form construction of local color-kinematics satisfying numerators in
the NLSM, we will discuss its applicability to all orders in $\ap$. 

It is worth mentioning a related \LitseyJFA\ approach to
constructing double-copy representations known as the Cachazo--He--Yuan (CHY)
formalism~\refs{\CHY, \CachazoHCA, \DPellis}, which generalizes the
four-dimensional connected prescription of Roiban, Spradlin, Volovich, and Witten
\refs{\WittenNN, \RoibanYF} to general dimensions.
Similar to string theory, scattering amplitudes in the
CHY framework are derived from punctured Riemann surfaces\foot{The CHY
integrands for gluon and graviton scattering have direct antecedents in the
heterotic string and the type-II superstring, respectively \refs{\GomezWZA, \BerkovitsXBA}. 
Also see ref.~\OSEym\ for a careful discussion of subtle differences between CHY
(tree-level) integrands for Einstein--Yang--Mills \ChyEym\ and correlation functions
 of the heterotic string.}.
Exploiting a CHY description of Yang--Mills theory and the
NLSM model, ref.~\CachazoNLSM\ offered the first double-copy realization of
self-dual Born--Infeld~\BI\ scattering amplitudes.   

The idea that there
can exist a duality between electric and magnetic field densities is as old as
gauge theory.  Satisfied by sourceless Maxwell electrodynamics, this natural
duality has inspired analyses and generalizations that have
been key to understanding aspects of supersymmetry, symmetry breaking, and string
theory, starting with perhaps most famously the Born--Infeld non-linear generalization of 
electromagnetics~\Schrodinger. The emergence of duality invariance in the form
of Born--Infeld  scattering due to a double-copy interplay between YM and the low-energy 
limit of abelian $Z$-theory is remarkable. 
In concordance with the structure of
open-string amplitudes given as a double copy between Yang--Mills constituents and
$Z$-theory disk-integrals \Ztheory, the double-copy representation of Born--Infeld
amplitudes as its surviving abelian low-energy limit serves as a key check that
our observation \preview\ holds to all multiplicities, beyond the explicit
verification at $n\leq 9$ points we report on here.

\subsec $Z$-theory amplitudes
\par\subseclab\sectZIntReview

Tree-level scattering amplitudes of open-string states are determined by iterated
integrals on the boundary of a disk worldsheet. Massless $n$-point amplitudes of
the open superstring \nptTreeI\ and conjecturally those of the open bosonic string
\HuangTAG\ possess cyclic integrands of the following form:
\eqn\Zintdef{
Z_{P}(q_1,q_2,\ldots,q_n) \equiv \ap^{n-3}\!\!\!\!
\int \limits_{D(P)} {\dz_1 \ \dz_2 \ \cdots  \ \dz_n \over {\rm vol}(SL(2,\Bbb R))}
 { \prod_{i<j}^n |z_{ij}|^{\ap s_{ij}}  \over z_{q_1 q_2} z_{q_2 q_3} \ldots z_{q_{n-1} q_n} z_{q_n q_1}} \, .
}
The universal and permutation-symmetric Koba-Nielsen factor $\prod_{i<j}^n
|z_{ij}|^{\ap s_{ij}}$ built from differences $z_{ij} \equiv z_i - z_j$ is
accompanied by a cyclic product of propagators $z_{q_i q_{i+1}}^{-1}$ indicated by
the labels $(q_1,q_2,\ldots,q_n)$ on the left hand side. The additional subscript 
$P \equiv p_1 p_2 \ldots p_n$ encodes the ordering for the iterated integrals,
\eqn\domain{
D(P) \equiv \{ (z_1,z_2,\ldots,z_n) \in \Bbb R^n\ | \ -\infty < z_{p_1} < z_{p_2} < \ldots < z_{p_n} < \infty \} \ ,
}
and thereby the accompanying Chan--Paton trace over gauge-group generators $t_{p_1}
t_{p_2}\ldots t_{p_n}$. The inverse volume of the
conformal Killing group of the disk instructs one to drop
any three variables of integration $z_i,z_j,z_k$ and
to compensate with a Jacobian $z_{ij}z_{ik} z_{jk}$, e.g.
\eqn\volsl{
\int \limits_{D(12\ldots n)}\!\!\!\!\!  {\dz_1 \ \dz_2 \ \cdots  \ \dz_n \over {\rm vol}(SL(2,\Bbb R))} = z_{1,n-1} z_{1,n} z_{n-1,n} \int \limits^{z_{n-1}}_{z_1} \dz_{n-2} \int\limits^{z_{n-2}}_{z_1} \dz_{n-3} \ldots \int\limits^{z_4}_{z_1} \dz_{3}\int\limits^{z_3}_{z_1} \dz_{2}  \ .
}
The unintegrated variables can then be fixed to any real values such as $(z_{1}, z_{{n-1}} ,  z_{n})=(0,1,\infty)$. 
Finally, the Mandelstam variables are defined in terms of lightlike momenta $k_i$:
\eqn\defmand{
s_{ij} \equiv k_i \cdot k_j,\quad s_{i_1 i_2\ldots i_p} \equiv
{1\over 2}(k_{i_1}+k_{ i_2}+\cdots +k_{i_p})^2 \ .
}
Their appearance in the open-superstring amplitudes leads us to view the integrals \Zintdef\ as defining the
tree-level S-matrix of $Z$-theory, the collection of putative scalar effective field theories 
that incorporate all the $\ap$-expansion on a disk worldsheet.

Inspired by progress in string-theory scattering
organization, it was realized in the 1990's that gauge-theory
amplitude calculations simplify tremendously by considering
ordered gauge invariants depending only upon kinematics -- 
called color-ordered or color-stripped partial amplitudes.  The full
color-dressed S-matrix elements could be obtained by summing over a
product of these color-ordered amplitudes with appropriate
color-weights, either somewhat redundantly in a trace basis, or more
efficiently in the Del Duca-Dixon-Maltoni basis of \KKLance.  The
advantages in considering stripped or ordered partial amplitudes are
enormous -- they grow exponentially rather than factorially in local
diagram contributions.  Doubly-ordered amplitudes like the \Zintdef\
represent a further generalization -- stripping out another ordering
when one can be defined.  One can recover from \Zintdef\ the more
familiar field-theory color-ordered amplitudes by summing over the
stringy $P$-orders taking products with Chan--Paton traces.

\subsubsec Symmetries
\par\subseclab\sectSymmetriesReview

For a fixed choice of the integration domain $D(P)$, the integrals \Zintdef\
associated with different permutations of $q_1,q_2,\ldots,q_n$ satisfy
the same relations as color-ordered YM amplitudes. Apart from the obvious cyclic
symmetry and reflection parity,
\eqnn\cycZ
\eqnn\refZ
$$\eqalignno{
Z_{P}(q_1,q_2,q_3 \ldots,q_n) &= Z_{P}(q_2,q_3, \ldots,q_n,q_1)  &\cycZ \cr
Z_{P}(q_1,q_2, \ldots,q_n) &= (-1)^n Z_{P}(q_n,\ldots, q_2,q_1)\,, &\refZ
}$$
partial fraction rearrangements of the integrand and integration-by-parts
relations can be written as \Ztheory,
\eqnn\ZKK
\eqnn\ZBCJ
$$\eqalignno{
0 &= Z_{P}(1,A,n,B) - (-1)^\len{B}\sum_{\s\in A\shuffle \tilde B}
Z_{P}(1,\sigma,n)\,,\quad\forall A,B &\ZKK\cr
0 &= \sum_{j=2}^{n-1} (k_{q_1}\cdot k_{q_2 q_3\ldots q_j}) Z_{P}( q_2,q_3,\ldots
,q_{j},q_1 ,q_{j+1}, \ldots ,q_n)\ , &\ZBCJ
}
$$
where $A=a_1a_2 \ldots a_{\len{A}}$ and $B=b_1 b_2 \ldots b_{\len{B}}$ represent
arbitrary sets of particle labels, and $\tilde B$ denotes
the transpose of the set $B$. Furthermore, the shuffle product is defined by~\ReutBook
\eqn\Shrecurs{
\emptyset\shuffle A = A\shuffle\emptyset = A,\qquad
A\shuffle B \equiv a_1(a_2 \ldots a_{|A|} \shuffle B) + b_1(b_2 \ldots b_{|B|}
\shuffle A)\,.
}
Note that \ZKK\ and \ZBCJ\ take exactly the same form as the KK
relations \KKsym\ and the BCJ relations \BCJ\ among
$A_{\rm YM}(q_1,q_2,\ldots,q_n)$ (see also \fundBCJ),
which are well-known to yield an $(n{-}3)!$-element
basis. Analogous relations among disk integrals \Zintdef\ with the same integrands
$Q=(q_1,q_2,\ldots,q_n)$ but different orders $P=p_1p_2\ldots p_n$ include
cyclicity and reflection
\eqn\cycref{
Z_{p_1p_2\ldots p_n}(Q) = Z_{p_2 p_3\ldots p_n p_1}(Q)
= (-1)^n Z_{p_n \ldots p_2 p_1}(Q) \ ,
}
and additional relations follow from monodromy properties of the worldsheet
\refs{\BjerrumBohrRD, \StiebergerHQ}
\eqn\PBCJ{
0 = \sum_{j=2}^{n-1} \exp\! \big[i\pi \alpha' (k_{p_1} \cdot k_{p_2 p_3 \ldots p_j}) \big]
Z_{p_2p_3\ldots p_{j}p_1 p_{j+1} \ldots p_n}(Q) \ ,
}
which also yield an $(n{-}3)!$ basis of integration domains.

These symmetry properties underpin our viewpoint on \Zintdef\ as the doubly-partial
amplitudes of $Z$-theory which by \ZKK\ and \ZBCJ\ satisfy the color-kinematics
duality in the integrand orderings to all orders in $\alpha'$. The
additional $\alpha'$-dependence in the relations \PBCJ\ among the
integration domain orderings, on the other hand, imprint the monodromy properties of the disk
worldsheets on the S-matrix of $Z$-theory upon summing over the product with 
relevant Chan--Paton factors.

\subsec The field-theory limit
\par\subseclab\sectwothree

In the field-theory limit $\alpha' \rightarrow 0$, the disk integrals \Zintdef\ yield
kinematic poles that correspond to
the propagators of cubic diagrams \refs{\scherk,\Frampton}.
As a convenient tool to describe the pole structure, we recall the theory
of a bi-adjoint scalar $\phi \equiv \phi_{a|b} t_a \otimes \tilde t_b$ with a cubic interaction
\eqn\lagphicube{
{\cal L}_{\rm bi-adjoint}= {1\over 2} \partial_\mu \phi_{a|b} \partial^\mu \phi_{a|b}
+ {1\over 3} f_{acg} \tilde f_{bdh} \phi_{a|b} \phi_{c|d} \phi_{g|h} \ .
}
Doubly-partial amplitudes $m[P|Q]$ are
defined to track the traces of gauge-group generators $t_a$ and $\tilde t_b$
in the tree amplitudes of the above scalar theory \DPellis,
\eqn\defDP{
M_{\phi^3} = \!  \! \! \! \sum_{\sigma,\rho \in S_{n-1}}
\! \! \! \! {\rm Tr}(t_1 t_{\sigma(2)} \ldots t_{\sigma(n)})
 {\rm Tr}(\tilde t_1 \tilde t_{\rho(2)}\ldots \tilde t_{\rho(n)})\,
m[1,\sigma(2,\ldots,n)|1,\rho(2,\ldots,n)]\,.
}
Following the all-multiplicity techniques
of \nptTreeII, the field-theory limits of disk integrals have been written in 
terms of doubly-partial amplitudes as \DPellis,
\eqn\ftlim{
\lim_{\ap\to0} Z_{P}(Q) = m[P|Q]\,,
}
identifying the bi-adjoint scalar theory \lagphicube\ as the
low-energy limit\foot{The factor of $(\ap)^{n-3}$ in the
definition \Zintdef\ of $Z_{P}(Q)$ guarantees that the leading term
in the low-energy expansion is of order $s_{ij}^{3-n}$, without any
accompanying factors of $\ap$.} of $Z$-theory, see \FTlimit\ for an
efficient Berends--Giele implementation of \ftlim.

\subsec Abelian limit

Recall that the color-dressed $n$-point
tree amplitude of the open superstring is given by
\eqn\Mstring{
M_{\rm open }(\ap) = \!\!\!\!\sum_{\s \in S_{n-1}}\!\!\!
{\rm Tr} \big[
\, t_{a_{1}} \, t_{a_{\s(2)}} \cdots\, t_{a_{\s(n)}} \, \big]
\,A_{\rm open}(1,\s(2,3,\ldots,n-1,n);\ap) \ ,
}
where the Chan--Paton-stripped amplitudes determined in \refs{\nptTreeI} were later
identified in \Ztheory\ to exhibit a KLT-like structure
\eqnn\Ngluon
$$\eqalignno{
A_{\rm open}&(1,\sigma(2,3,\ldots,n);\ap) = \! \! \sum_{\rho,\tau \in S_{n-3}}
\! \! Z_{1\sigma(2,3,\ldots ,n)}(1,\rho(2,3,\ldots,n-2),n,n-1) &\Ngluon \cr
& \ \ \ \ \ \times S[ \rho(23\ldots n-2) | \tau(23\ldots n-2) ]_1
A_{\rm YM}(1,\tau(2,3,\ldots,n-2),n-1,n)\,.
}$$
The symmetric matrix $S[\rho|\tau]_1$ in \Ngluon\ encodes
the field-theory limit ($\ap\to0$) of KLT relations~\KLT\ to all multiplicities \BernKLT\ and 
admits the following recursive definition\foot{
The field-theory KLT matrix was originally defined in non-symmetric form
in \BernKLT, later rewritten in \refs{\BohrMomKer,\fundBCJ} with
the symmetric form used in \Ztheory.
Inspired by equation~(3.8) of \DuTBC\ we arrived at the novel
recursive definition \mkerA, which generalizes to all orders
in $\ap$ in an obvious manner \MomKer.},
\eqn\mkerA{
S[A,j|B,j,C]_i =
(k_{iB}\cdot k_{j}) S[A|B,C]_i,
\qquad S[\emptyset|\emptyset]_i \equiv 1\,,
}
where $A$, $B$ and $C$ are arbitrary multiparticle labels such that $\len{A} =
\len{B}+\len{C}$ and the multiparticle momentum is defined by
$k_{iB}\equiv k_i + k_{b_1} + \cdots + k_{b_\len{B}}$.
For example $S[2,3,4|2,4,3]_1 = (k_{12}\cdot k_4)S[2,3|2,3]_1
= (k_{12}\cdot k_4)(k_{12}\cdot k_3)S[2|2]_1
= (k_{12}\cdot k_4)(k_{12}\cdot k_3)(k_1\cdot k_2)$. The doubly-partial amplitudes \defDP\ of bi-adjoint scalars,
more precisely the $(n{-}3)! \times (n{-}3)!$-basis $m[1,\rho(2,\ldots,n{-}2),n,n{-}1|1,\tau(2,\ldots,n{-}2),n{-}1,n]$, furnish the inverse of this matrix \DPellis.   

Note that the Chan--Paton-ordering $\sigma$ of the string amplitude
in \Ngluon\ enters globally as the integration domain of the
$Z_{1\sigma(2,3,\ldots, n)}(\ldots)$ and does not interfere with the
permutation sums over $\rho$ and $\tau$ in \Ngluon. Accordingly, its
specialization to abelian open superstrings is obtained by setting
all the Chan--Paton traces to unity and yields,
\eqnn\Nphoton
$$\eqalignno{
M_{\rm abelian}(\ap) & = \! \! \sum_{\rho,\tau \in S_{n-3}} \! \!
Z_{\times }(1,\rho(2,3,\ldots,n-2),n,n-1) &\Nphoton \cr
& \ \ \ \ \ \times S[ \rho(23\ldots n-2) | \tau(23\ldots n-2) ]_1 A_{\rm
YM}(1,\tau(2,3,\ldots,n-2),n-1,n)\,,
}$$
where
\eqn\abdisk{
Z_{\times}(q_1,q_2,\ldots,q_n) \equiv
\sum_{\sigma \in S_{n-1}} Z_{1\sigma(2,3,\ldots,n)}(q_1,q_2,\ldots,q_n)
}
defines the {\it abelian disk integrals} or the partial amplitudes of abelian $Z$-theory
whose $\ap$-expansion will be discussed below.  

\subsec $\alpha'$-expansion
\par\subseclab\sectwofive

The $\alpha'$-expansion of the disk integrals \Zintdef\ gives rise
to multiple zeta values (MZVs),
\eqn\MZVdef{
\zeta_{n_1,n_2,\ldots, n_r} \equiv \sum_{0<k_1<k_2<\ldots <k_r}^{\infty} k_1^{-n_1} k_2^{-n_2} \ldots k_r^{-n_r} 
\ , \ \ \ \ n_r \geq 2 \ ,
}
which are characterized by their weight $w=n_1+n_2+\ldots+n_r$ and depth $r$. 
More precisely, the order $(\alpha')^w$ of the disk integrals in \Zintdef\ is accompanied
by products of MZVs with total weight $w$ (where the weight is understood to be
additive in products of MZVs); a property known as {\it uniform transcendentality}.
This has been discussed in the literature of both
mathematics \mathMZV\ and physics \refs{\nptTreeII, \StiebergerRR, \SchlottererNY}
and can for instance be proven by the recursive construction\foot{At multiplicities
five, six and seven, explicit results for the leading orders are available for
download on \WWW, along with the building blocks for eight and nine points.} of
disk integrals using the Drinfeld associator \BroedelAZA.

The combination of all integration orders to obtain abelian disk integrals projects
out a variety of MZVs from the $\alpha'$-expansion of \abdisk. As elaborated in
section \SecAllOrderSystematics, these cancellations include the field-theory limit
\ftlim\ and the coefficients of odd Riemann zeta values $\zeta_{2k+1}$ 
without accompanying factors of $\zeta_{2n}$. Moreover, 
abelian disk integrals of odd multiplicity vanish at all orders in $\alpha'$ by the
reflection property \cycref,
\eqn\noodd{
Z_{\times}(q_1,q_2,\ldots,q_{2k+1})=0 \ .
}
It turns out that the leading low-energy contribution to abelian disk integrals
of even multiplicity $n$ arises from the order $\alpha'^{n-2}$ and stems solely
from the even Riemann zeta values such as ($B_{2k}$ are the Bernoulli numbers)
\eqn\zetaeven{
\zeta_2 = {\pi^2 \over 6}\,,\qquad
\zeta_4 = {\pi^4 \over 90}\,,\qquad
\zeta_6 = {\pi^6 \over 945}\,,\qquad\ldots\qquad
\zeta_{2k} = (-1)^{k-1} { (2\pi)^{2k} B_{2k} \over 2 (2k)!}\,.
}
The impact of these selection principles on the $\alpha'$-expansion of abelian
$Z$-theory in connection with NLSM amplitudes will be explored in section~\secthree.
In light of the ubiquitous appearance of MZVs in both abelian and non-abelian
$Z$-theory, one might be tempted to derive the capital letter in the theory's name
from ``zeta''.

\newsec NLSM amplitudes from string theory
\par\seclab\sectwofour

\noindent Although the superstring spectrum does not include any bi-adjoint
scalar,
 the doubly-partial amplitudes
\defDP\ emerge naturally from the low-energy limit of the $Z$-theory amplitudes
  \Zintdef\ contributing to the open string.  Perhaps more familiarly,
  the color-stripped amplitudes of the bi-adjoint scalar appear as the $\ap\to0$ limit
  of the Chan--Paton dressed doubly-partial $Z$-theory amplitudes.
 In this work, we show that
the NLSM tree-level amplitudes can be obtained from the abelian disk integrals
\abdisk.

To see this, note that the Born--Infeld action emerges as the leading low-energy
contribution to abelian amplitudes in supersymmetric
string theory~\BIref.
Therefore, the expression for $M_{\rm abelian}(\ap) $ on the right-hand side of
\Nphoton\ must reduce to the Born--Infeld amplitude whose
KLT-like double-copy structure has recently been identified by Cachazo, He and Yuan
\CachazoNLSM,
\eqnn\BIamp
$$\eqalignno{
M_{\rm BI}& = \! \! \sum_{\rho,\tau \in S_{n-3}} \! \! A_{\rm NLSM}(1,\rho(2,3,\ldots,n-2),n,n-1) &\BIamp \cr
& \ \ \ \ \ \times S[ \rho(23\ldots n-2) | \tau(23\ldots n-2) ]_1 A_{\rm YM}(1,\tau(2,3,\ldots,n-2),n-1,n) \ .
}$$
Comparing \Nphoton\ with \BIamp\ and assuming linear independence 
of the YM partial amplitudes in the BCJ basis leads to the conclusion
that the abelian
$Z$-amplitudes \abdisk\ reduce to color-ordered NLSM tree amplitudes at low
energies, 
\eqn\NLSMamp{
A_{\rm NLSM}(1,2,\ldots,n) =
\lim_{\alpha' \rightarrow 0} (\alpha')^{2-n}
Z_{\times}(1,2,\ldots,n)\,.
}
This limit is non-singular due to the cancellation of low-energy
orders below $(\alpha')^{n-2}$ in {\it abelian} $n$-point
$Z$-amplitudes, see section \SecAllOrderSystematics\ for further
details.
The emergence of NLSM amplitudes in \NLSMamp\ has been explicitly verified 
up to $n=9$, using the expansion method of
\BroedelAZA\ to probe the $\alpha'^6$-order at the highest non-trivial multiplicity
$n=8$. As an immediate consistency condition for the validity of \NLSMamp, note
that the KK and BCJ relations satisfied by the NLSM amplitudes \chendu\ correspond
to the following identities of the abelian integrals,
\eqnn\ABKK
$$\eqalignno{
0 &= Z_\times(1,A,n-1,B) - (-1)^\len{B}\sum_{\sigma\in A\shuffle \tilde B}
Z_\times(1,\sigma,n-1)\,,\quad\forall A,B &\ABKK\cr
0 &= \sum_{j=2}^{n-1} (k_{q_1}\cdot k_{q_2 q_3\ldots q_j})
Z_\times( q_2,q_3,\ldots,q_{j},q_1 ,q_{j+1}, \ldots ,q_n) \ ,
}
$$
which are a consequence of \ZKK\ and \ZBCJ.

Besides reproducing NLSM amplitudes, higher $\alpha'$-orders of abelian disk
integrals \abdisk\ yield natural higher-mass dimension extensions of the NLSM which
will all satisfy KK and BCJ relations \ABKK. More precisely, the symmetry
properties \ABKK\ hold separately at each order in $\alpha'$, and in fact for the
coefficients of any MZV which is conjecturally linearly independent over $\Bbb Q$.
Hence, abelian disk integrals can be viewed as a factory for effective theories
with any number of derivatives, and each such theory obeys the duality between color and
kinematics. The discussion of these $\alpha'$-corrections to the NLSM will be the
main focus of section \secthree.

\newsec Higher-derivative corrections to the NLSM
\par\seclab\secthree

\subsec Four points

At four points, monodromy relations \PBCJ\ \refs{\BjerrumBohrRD, \StiebergerHQ} allow
to compactly
express the abelian disk integral \abdisk\ in terms of any $Z_{P}(q_1,q_2,q_3,q_4) $, e.g.
\eqnn\combfour
$$\eqalignno{
Z_{\times}(1,2,4,3) &= 2 \Big( 1 + { \sin (\alpha' \pi s_{23}) \over  \sin (\alpha' \pi s_{13}) }
+ { \sin (\alpha' \pi s_{12}) \over  \sin (\alpha' \pi s_{13}) }\Big)
Z_{1234}(1,2,4,3) \ . &\combfour\cr
}$$
It is straightforward to see using the form of the Veneziano amplitude
\eqn\veneziano{
s_{12}Z_{1234}(1,2,4,3) = {\Gamma(1+\ap s_{12})\Gamma(1+\ap s_{23})\over
\Gamma(1+\ap (s_{12} + s_{23}))}
}
together with the identities
\eqn\idsrew{
\sin(\pi x) = {\pi\over \Gamma(1-x)\Gamma(x)},\quad
\ln(\Gamma(1+x)) = -\gamma x + \sum_{k=2}^\infty {\zeta_k\over k}(-x)^k
}
that the four-point abelian integral \combfour\ can be written as
\eqn\fourclosed{
Z_\times(1,2,4,3) =  {2 \big[ \sin(\pi \alpha' s_{12}) + {\rm cyc}(1,2,3) \big]
\over \pi \alpha' s_{12} s_{13} } \, \exp
\Big( \sum_{k=2}^{\infty} \frac{ \zeta_{k}}{k}(-\alpha')^{k} \big[
s_{12}^{k} +s_{23}^{k} + s_{13}^{k} \big] \Big)\,.
}
The abelian integral \fourclosed\ not only reproduces the standard
four-point NLSM amplitude $A_{\rm NLSM}(1,2,3,4) = - \pi^2 s_{13}$
at its lowest $\ap$ order
(note the swap of legs $3\leftrightarrow 4$),
but also implies an infinite series of higher-derivative corrections \deRooXV,
\eqnn\fourab
$$\eqalignno{
&Z_{\times}(1,2,3,4) =  - \alpha'^2\pi^2 s_{13} \times \Big(1 + \frac{1}{2} \zeta_2 \sigma_2 + \zeta_3 \sigma_3 + \frac{3}{10} \zeta_2^2 \sigma_2^2 + (\zeta_5 + \frac{1}{2}\zeta_3 \zeta_2) \sigma_2 \sigma_3  \cr
& \ \ \ \ \ \ \ \ \ \ \ \ + \frac{1}{2} \zeta_3^2 \sigma_3^2 + \frac{\zeta_2^3}{280} (31 \sigma_3^2 + 51 \sigma_2^3)
+(\zeta_7 + \frac{1}{2} \zeta_5 \zeta_2 + \frac{3}{10} \zeta_3 \zeta_2^2) \sigma_2^2 \sigma_3 &\fourab \cr
& \ \ \ \ \ \ \ \ \ \ \ \ +(\zeta_3 \zeta_5+ \frac{1}{4} \zeta_2 \zeta_3^2 ) \sigma_2 \sigma_3^2 + \frac{ \zeta_2^4 \sigma_2}{1400}(67 \sigma_3^2 + 31 \sigma_2^3)  +\ldots \Big) \ , 
}$$
where we defined $\sigma_2 \equiv {1 \over 2} \ap^2 (s_{12}^2 + s_{13}^2 + s_{23}^2)$
and $\sigma_3 \equiv - \ap^3 s_{12}s_{23} s_{13}$.
Note that the terms inside parenthesis in \fourab\ are invariant
under permutations, thereby manifesting the BCJ and KK relations \ZKK\ obeyed
by $Z_\times(1,2,3,4)$.

\subsec Six points
\par\subseclab\secthreetwo

The $\alpha'$-expansion of six-point disk integrals \Zintdef\ was pioneered in \refs{\OprisaWU,\multigluon}
and later on aligned into systematic all-multiplicity methods using polylogarithms
\Ztheory\ or the Drinfeld associator \BroedelAZA\ (see also \DrummondVZ).
When summing over the $5!$ integration domains to obtain an abelian six-point disk integral \abdisk,
the leading $\alpha'$-orders associated with $s_{ij}^{-3}$,
$\alpha'^2\zeta_2s_{ij}^{-1}$ and $\alpha'^3\zeta_3$ turn out to cancel, see
section \SecAllOrderSystematics\ for further details. The first non-vanishing order $\sim
\alpha'^4 \zeta_4$ coincides with the six-point NLSM amplitude \NLSMsix,
\eqnn\confirmsix
$$\eqalignno{
Z_{\times}(1,2,3,4,5,6) &= \alpha'^4 \pi^4 \Big\{ s_{12}  - \frac{ (s_{12}+s_{23}) (s_{45}+s_{56}) }{2 s_{123}} + {\rm cyc}(1,2,3,4,5,6) \Big\} + {\cal O}(\alpha'^6) 
\cr
&= \alpha'^4 A_{\rm NLSM}(1,2,3,4,5,6) +  {\cal O}(\alpha'^6) \ , &\confirmsix
}
$$
in agreement with the general claim \NLSMamp. Beyond the order $\alpha'^4\zeta_4$
of the NLSM, an infinite tower of corrections occurs in the expansion of
$Z_{\times}(1,2,\ldots,6)$, starting with $\alpha'^6\zeta_6$,
$\alpha'^7\zeta_4\zeta_3$, $\alpha'^8\zeta_8$, $\alpha'^9\zeta_6 \zeta_3$ and
$\alpha'^9\zeta_4 \zeta_5$. The lowest-order corrections are given by
\eqnn\sigsixtwo
$$\eqalignno{
Z_{\times}(1,&{}2,3,4,5,6) \, \big|_{\alpha'^6}=
\frac{\pi^6}{12} \Bigl[
-\frac{(s_{12} + s_{23}) (s_{12}^2 + s_{12} s_{23} + s_{23}^2) (s_{45} + s_{56})}{s_{123}}
+4 s_{12} s_{23} s_{234} \cr
&
+ 4 s_{12} s_{23} s_{345} - 4 s_{12} s_{23} s_{34} + 2 s_{12} s_{23} s_{56} + 
 2 s_{12} s_{23} s_{45} + 2 s_{12} s_{34} s_{123} + 2 s_{12} s_{34} s_{234} \cr
&
+ s_{12} s_{34} s_{345} +
 s_{12}^3 + 2 s_{12}^2 s_{45} + 2 s_{12}^2 s_{234} - 2 s_{12} s_{234}^2
 - 4 s_{12} s_{123} s_{234} -
 2 s_{23} s_{123} s_{234}\cr
&
- 4 s_{34} s_{123} s_{234}-
 \frac{1}{2} s_{12} s_{45} s_{123} - \frac{1}{2} s_{12} s_{45} s_{345} + s_{123}^2 s_{234} +
 s_{123} s_{234}^2 + \frac{1}{3} s_{12} s_{34} s_{56} \cr
& + \frac{4}{3} s_{123} s_{234} s_{345} + {\rm cyc}(1,2,3,4,5,6) \Bigr]  \ ,&\sigsixtwo
}$$
and the expression for the terms of order $\ap^7\zeta_4\zeta_3$ can
be found in \sigsixthree. For both of them, the residue of the
kinematic pole in $s_{123}^{-1}$ is easily seen to factorize
correctly on the relevant orders in the four-point $\ap$-expansion
\fourab.

\subsec All order-systematics
\par\subseclab\SecAllOrderSystematics

In order to discuss the all-multiplicity systematics of the $\alpha'$-expansion of abelian disk integrals,
we recall the patterns of MZVs in open-string amplitudes identified in \SchlottererNY.
A particularly convenient basis for that purpose is furnished by the $(n-3)! \times (n-3)!$ integrals\foot{Note that in a frame where $(z_1,z_{n-1},z_n)=(0,1,\infty)$, the integrals in \KLTopen\ take the form \nptTreeI
%
$$\eqalignno{
F_\Sigma{}^{\rho} &= (-\ap)^{n-3} \! \! \! \! \! \! \! \!  \! \! \! \!  \int\limits_{0\leq z_{\Sigma(2)} \leq z_{\Sigma(3)} \leq \ldots \leq z_{\Sigma(n-2)} \leq 1}   \! \! \! \! \! \! \! \! \! \! \! \!   \dd z_2 \, \dd z_3\, \ldots \, \dd z_{n-2} \prod_{i<j}^{n-1} |z_{ij}|^{\alpha' s_{ij}} \, { s_{1 \rho(2)} \over z_{1 \rho(2)}}
\left(  { s_{1 \rho(3)} \over z_{1 \rho(3)}} + { s_{\rho(2) \rho(3)} \over z_{\rho(2) \rho(3)}}
\right) 
\cr 
& \ \ \ \ \ \times \left(  { s_{1 \rho(4)} \over z_{1 \rho(4)}} + { s_{\rho(2) \rho(4)} \over z_{\rho(2) \rho(4)}} + { s_{\rho(3) \rho(4)} \over z_{\rho(3) \rho(4)}}
\right) \ldots \left(  { s_{1 \rho({n-2})} \over z_{1 \rho({n-2})}} + { s_{\rho(2) \rho({n-2})} \over z_{\rho(2) \rho({n-2})}} +\ldots + { s_{\rho({n-3}) \rho({n-2} )} \over z_{\rho({n-3}) \rho({n-2}) }}
\right) \ .
}$$
}
appearing in the $n$-point amplitude \Ngluon\ of the
superstring \refs{\nptTreeI, \Ztheory}
\eqn\KLTopen{
F_\Sigma{}^{\rho} \equiv  \!\! \sum_{\tau \in S_{n-3}} \!\!S[\rho(23\ldots n-2) | \tau(23\ldots n-2)]_1 Z_{1,\Sigma(23\ldots n-2),n-1,n}(1,\tau(2,3,\ldots,n-2),n,n-1)  \ .
}
These integrals form a square matrix indexed by integration
domains $\Sigma$ and integrands $\rho$, and the multiplication
with the KLT matrix $S[\cdot|\cdot]_1$ defined in \mkerA\ ensures
that all the entries are analytic in $\alpha'$, i.e.\ that there
are no poles in any $s_{i_1\ldots i_p}$. The pattern of MZVs in
the power-series expansion is based on matrix multiplications \SchlottererNY:
\eqnn\Fexpa
$$\eqalignno{
F &= (1 + \zeta_2 P_2 +  \zeta_2^2 P_4 +  \zeta_2^3 P_6 + \zeta_2^4 P_8 +  \ldots) 
&\Fexpa \cr
&\times \Big( 1 + \zeta_3 M_3 + \zeta_5 M_5 +\frac{1}{2} \zeta_3^2 M_3^2 + \zeta_7 M_7 + \zeta_3 \zeta_5 M_5 M_3+ \frac{1}{5} \zeta_{3,5} [M_5,M_3] + \ldots \Big) \ .
}$$
Both $P_{w}$ and $M_{w}$ denote $(n-3)!\times (n-3)!$ matrices whose entries are
degree-$w$ polynomials in $\alpha' s_{ij}$ with rational coefficients. The explicit form of
these entries can be determined from polylogarithm manipulations
\Ztheory\ or the Drinfeld associator \BroedelAZA, and examples at multiplicity
$n\leq 7$ are available for download from \WWW. As a first non-trivial statement
of \Fexpa, for instance, the coefficient of $\zeta_2 \zeta_3$ is given by the
matrix product $P_2 M_3$ combining the constituents at the $\zeta_2$- and
$\zeta_3$-orders of $F$.

\subsubsec Selection rule for the zero'th order in $\zeta_2$

The monodromy relations \PBCJ\ among different color-orderings of $A_{\rm
open}(\ldots)$ can be viewed as deformation of the BCJ relations by even powers of
$\alpha' \pi s_{ij}$, i.e.\ by $\zeta_{2k}$ according to \zetaeven. These
$\alpha'$-corrections only interfere with the left-multiplicative factors of
$\zeta_{2k}P_{2k}$ in the first line of \Fexpa.
Therefore
the entire second line of \Fexpa\ -- in
fact any product of matrices $M_{2k+1}$ -- preserves the BCJ and KK
relations\foot{This argument firstly appeared in the discussion of BCJ relations
among amplitudes from higher-mass dimension operators \BroedelRC, and a similar
statement in the context of the heterotic string can be found in \StiebergerHBA.}
for $(M_{2k_1+1} \ldots M_{2k_n+1}) A_{\rm YM}$.

Accordingly, the disk integrals' coefficient of $\zeta_{2k+1}$, $\zeta_{3}
\zeta_{5}$, $\zeta_{3,5}$ as well as suitable generalizations at higher weight and
depth \refs{\BlumleinCF, \SchlottererNY}\foot{The choice of MZVs at a given weight
to represent the $\ap$-expansion of disk integrals is ambiguous, and we will follow the conventions of
\refs{\BlumleinCF, \SchlottererNY} to take $\{\zeta_8, \, \zeta_3 \zeta_5 , \,
\zeta_2 \zeta_3^2 , \, \zeta_{3,5}\}$ as the conjectural $\Bbb Q$-basis of
weight-eight MZVs. Different choices lead to redefinitions of the matrices $P_w,
M_w$, e.g.\ $P_8$ is shifted by a rational multiple of $[M_3,M_5]$ when trading
$\zeta_{3,5}$ for another MZV of depth~$\geq 2$.} satisfy the BCJ and KK
relations. Once we collectively denote any MZV in the second line of \Fexpa\ by
$\zeta_M \in \{\zeta_{2k+1}, \zeta_{3} \zeta_{5},\zeta_{3,5},\ldots \}$, this can
be written as
\eqn\zbcj{
 \sum_{j=2}^{n-1} (k_{p_1} \cdot k_{p_2 p_3\ldots p_j})
  Z_{p_2p_3\ldots p_{j}p_1 p_{j+1} \ldots p_n}(q_1,q_2,\ldots,q_n)  \, \big|_{\zeta_{M}} = 0 \ .
}
Hence, these MZVs drop out from abelian disk integrals,
\eqn\selectzx{
Z_{\times}(q_1,q_2,\ldots,q_n)  \, \big|_{\zeta_{M}}  = 0 \ .
}

\subsubsec Selection rule for the first order in $\zeta_2$

Similarly, the $\alpha'$-deformed BCJ relations of $P_{2k} A_{\rm YM}$ encoded in
the monodromy relations directly carry over to the matrix products $P_{2k} M_{2\ell_1+1}
\ldots M_{2\ell_m+1} A_{\rm YM}$ with $\ell_j ,m\in \Bbb N$. This has a direct
implication for abelian disk integrals:
Whenever the coefficient of $\zeta_{2}^k$
vanishes by the monodromy relations to that order, the same vanishing statement
applies to all products of $\zeta_2^k$ with the entire second line $\sim \zeta_M$ of \Fexpa,
\eqn\moreselectzx{
Z_{\times}(q_1,q_2,\ldots,q_n)  \, \big|_{\zeta_{2}^k} = 0 \ \ \ \Rightarrow \ \ \
Z_{\times}(q_1,q_2,\ldots,q_n)  \, \big|_{\zeta_{2}^k \, \zeta_M} = 0\,.
}
The ``KK-like'' relations among the $\zeta_2$-orders of open superstring amplitudes \refs{\multigluon, \BjerrumBohrXE, \MafraKH} 
 for instance are known to annihilate permutation sums at multiplicities $n\geq 5$
 and therefore
\eqn\selectone{
Z_{\times}(q_1,q_2,\ldots,q_n)  \, \big|_{\zeta_{2} \, \zeta_{M}} = 0,
\qquad \forall \ n\geq 5\,.
}

\subsubsec Selection rule for higher orders in $\zeta_2$

Although the symmetry patterns associated with the $\zeta_4, \zeta_6,\ldots$-orders of open superstring 
amplitudes have not yet been studied, there is an indirect argument
to extend the selection rule \selectone\ to higher orders: Since the
low-energy limit of the abelian amplitude \Nphoton\ is known to stem from the
Born--Infeld action \BIref, the $n$-point amplitude cannot have contributions of orders below $\alpha'^{n-2}$.
In particular, this implies
\eqn\selecttwo{
Z_{\times}(q_1,q_2,\ldots,q_n)  \, \big|_{\zeta_{2}^k} = 0, \qquad
\forall \ k < {n\over2}-1 
}
and leads to an infinity of additional vanishing statements by \moreselectzx,
\eqn\selectthree{
 Z_{\times}(q_1,q_2,\ldots,q_n)  \, \big|_{\zeta_{2}^k \, \zeta_{M}} = 0 \ ,
 \qquad \forall \ k<{n\over2}-1 \ ,
}
with $\zeta_{ M }$ again referring to any MZV in the second line of \Fexpa.

Examples of the above selection rules on the low-energy regime of abelian disk integrals are
summarized in the subsequent table:
\bigskip
\def\dvrule{\vrule\hskip 0.02cm \vrule}
\def\hquad{\hskip0.63em\relax}
\vbox{
\halign{\strut\vrule\hfil\hquad $#$\hquad\hfil 
&\dvrule\hfil\hquad$#$\hquad\hfil &\dvrule\hfil\hquad$#$\hquad\hfil &\dvrule\hfil\hquad$#$\hquad\hfil
&\dvrule\hfil\hquad$#$\hquad\hfil &\vrule\hfil\hquad $#$\hquad\hfil
&\dvrule\hfil\hquad $#$\hquad\hfil &\vrule\hfil\hquad $#$\hquad\hfil
&\dvrule\hfil\hquad $#$\hquad\hfil &\vrule\hfil\hquad $#$\hquad\hfil &\vrule\hfil\hquad $#$\hquad\hfil
&\dvrule\hfil\hquad $#$\hquad\hfil &\vrule\hfil\hquad $#$\hquad\hfil &\vrule\hfil\hquad $#$\hquad\hfil & \vrule\hfil\hquad
 $#$\hquad\hfil \vrule \tabskip=0pt\cr
\noalign{\hrule}
n & \zeta_2  & \zeta_3  & \zeta_4 
& \zeta_5  & \zeta_2 \zeta_3 
& \zeta_6  & \zeta_3^2 
& \zeta_7  & \zeta_2 \zeta_5  & \zeta_4 \zeta_3 
& \zeta_8  & \zeta_3 \zeta_5  & \zeta_{3,5}  & \zeta_2 \zeta_3^2  \cr
\noalign{\hrule}
4 &  \checkmark  & \times  & \checkmark  & \times  & \checkmark 
& \checkmark  & \times 
& \times  & \checkmark  & \checkmark 
& \checkmark  & \times  & \times  & \checkmark 
 \cr 
\noalign{\hrule}
6 & \times  & \times  & \checkmark  & \times  & \times 
& \checkmark  & \times 
& \times  & \times  & \checkmark 
& \checkmark  & \times  & \times  & \times  \cr
\noalign{\hrule}
8 &  \times  & \times  & \times  & \times  & \times 
& \checkmark  & \times 
& \times  & \times  & \times 
& \checkmark  & \times  & \times  & \times  \cr
\noalign{\hrule}
10 &  \times  & \times  & \times  & \times  & \times 
& \times  & \times 
& \times  & \times  & \times 
& \checkmark  & \times  & \times  & \times  \cr
\noalign{\hrule}
}
\medskip\smallskip
{\leftskip=0pt\rightskip=20pt\noindent\ninepoint
{\bf Table 1.} Overview of the MZVs of weight $w \leq 8$ present in abelian disk integrals at
multiplicities $n=4,6,8,10$. In each of the fields marked by $\times$, the
selection rules \selecttwo\ and \selectthree\ forbid the appearance of
the respective MZV.}
}

These selection rules have been explicitly verified up to $\ap^7$.
By \abdisk\ this requires the $\ap$-expansion of the $Z$-integrals
\Zintdef\ for various orderings of the integration region. An
efficient algorithm to perform this task has been subsequently developed\foot{Earlier
versions of this manuscript described the rather challenging procedure originally used 
to extract these $\ap$-expansions.} in \BGap\ using a technique akin
to the Berends--Giele recursion for computing tree-level Yang--Mills
amplitudes \BerendsME. These $\ap$-expansions are 
readily available from the implementation in \gitrepo\ and all the results tested up to $\ap^7$
are compatible with the above selection rules.  Furthermore, advanced
consideration of handling the Chan--Paton dressing presented in \SAbZ\
reduces the order of $\ap$ that needs be extracted from $Z$-functions in 
abelianized calculations by $n{-}2$.

\subsec Simplifications in the odd zeta sector
\par\subseclab\secthreefour

It turns out that the laborious procedure to determine the $\alpha'$-expansion of
$Z_{\times}(\ldots)$ can be bypassed for all the $M_w$ matrices. Once we have
determined the contributions of the type $\zeta_{2k}P_{2k}$ from the first line of
\Fexpa,
\eqn\Zeven{
Z_{\times}^{{\rm even}}(q_1,q_2,\ldots,q_n)
\equiv
Z_{\times}(q_1,q_2,\ldots,q_n) \, \big|_{\zeta_M \rightarrow 0} \ ,
}
then the coefficient of $\zeta_{2k+1}$ or any other MZV in the second line of \Fexpa\
can be inferred by matrix multiplication
\eqnn\Zexpa
$$\eqalignno{
&Z_{\times}(1,\tau(2,3,\ldots,n-2),n-1,n) =  \sum_{\sigma \in S_{n-3}} 
\Big( 1 + \zeta_3 M_3 + \zeta_5 M_5 +\frac{1}{2} \zeta_3^2 M_3^2 &\Zexpa  \cr
&\ \ \ \ \ + \zeta_7 M_7 + \zeta_3 \zeta_5 M_3 M_5- \frac{1}{5} \zeta_{3,5} [M_5,M_3] + \ldots \Big)_\tau{}^{\sigma} Z_{\times}^{{\rm even}}(1,\sigma(2,\ldots,n-2),n-1,n) \ .
}$$
Note, however, that the multiplication order of $M_{w}$ matrices is reversed in \Zexpa\ as 
compared to \Fexpa. As before, matrix multiplication with any sequence of $M_{2k+1}$ 
propagates the BCJ and KK relations of $ Z_{\times}^{{\rm even}}$ to the
full integral $ Z_{\times}$.

One might wonder if the structure in \Zexpa\ can be refined and if the appearance of 
any $\zeta_{2k}$ in $Z_{\times}$ can be captured by combining a BCJ basis of $ A_{{\rm
NLSM}}(1,\sigma(2,\ldots,n{-}2),n{-}1,n)$ with polynomials in Mandelstam variables.
When insisting on local coefficients for the basis of NLSM amplitudes, this
scenario can be ruled out from a simple six-point example: In an ansatz of the form
\eqn\mtwo{
Z_{\times}(1,\tau(2,3,4),5,6) \, \big|_{\zeta_6} = \alpha'^4 \sum_{\sigma \in S_3} (M_2)_\tau{}^{\sigma}A_{{\rm NLSM}}(1,\sigma(2,3,4),5,6) \ ,
}
with the left-hand side given by \sigsixtwo, the entries of the $6\times 6$ matrix $M_2$ cannot be
chosen as degree-two polynomials in $\alpha' s_{ij}$. Hence, there is no {\rm
local} degree-two counterpart $M_2$ of the $M_{2k+1}$ matrices at six-points which preserves the
BCJ and KK relations.

\newsec Color-kinematic satisfying numerators
\par\seclab\secfour

\noindent As we will review, the fact that color-stripped NLSM amplitudes satisfy the 
BCJ relations ensures~\refs{\BCJfromKLT, \MomKer}\ that they admit a
color-kinematic satisfying representation at tree-level by virtue of the existence
of the KLT decomposition.  The fact that this holds to all multiplicity suggests
that the integrands of these theories, effective though they are, should also admit
color-kinematic satisfying numerators.  This intriguing possibility motivates
exploring what various closed forms for color-kinematic satisfying tree-level
numerators can be found.

The BCJ relations \ABKK\ among abelian disk integrals hold
separately at each order in $\alpha'$, more precisely for the
coefficients of all the MZVs which are conjecturally linearly
independent over $\Bbb Q$. Following the original derivation of BCJ
relations for YM amplitudes from the duality between color and
kinematics \BCJ, one should expect each MZV coefficient of
$Z_{\times}(\ldots)$ to admit a cubic-graph organization\foot{ In
spite of the Feynman vertices of valence $\geq 4$ in the NLSM
Lagrangian \Lag, one can always achieve a cubic-graph organization
of its scattering amplitudes by introducing targeted propagators $1
= {(k_i+k_j)^2 \over (k_i+k_j)^2}$ whose channels $i,j$ are lined up
with the accompanying color factors. This kind of bookkeeping has
been originally applied to the quartic vertex of Yang--Mills
theories \BCJ\ and can be straightforwardly extended to deformations
of the NLSM by vertices of arbitrary valence and order in
derivatives. }, where the $s_{ij}$-dependent numerators satisfy
kinematic Jacobi relations.  The latter apply to any triplet of
cubic diagrams whose color factors under a generic gauge group
(obtained from dressing each vertex by a structure constant
$f^{abc}$ of a non-abelian gauge group) sum to zero by the Jacobi
identity $f^{a_1 a_2 b} f^{b a_3 a_4} + {\rm cyc}(a_1,a_2,a_3)$, see
\figBCJ.

\ifig\figBCJ{The Jacobi identity $f^{a_1 a_2 b} f^{b a_3 a_4} + {\rm cyc}(a_1,a_2,a_3)$ implies the vanishing of 
 the color factors $C_i$, $C_j$ and $C_k$ associated to triplets of cubic graphs. In the above diagrams, the 
 legs $a_1,\ldots ,a_4$ may represent arbitrary cubic tree-level subdiagrams. The duality between color and kinematics 
 states that their corresponding kinematic numerators $N_i$ built from polynomials of Mandelstam invariant for the cases 
 of interest can be chosen such that $N_i+N_j+N_k=0$ whenever $C_i+C_j+C_k=0$ \BCJ.}
{\hbox{{\epsfxsize=0.20\hsize\epsfbox{figBCJ.1}}\raise26pt\hbox{$\ + \ $}%
{\epsfxsize=0.20\hsize\epsfbox{figBCJ.2}}\raise26pt\hbox{$\ + \ $}%
{\epsfxsize=0.20\hsize\epsfbox{figBCJ.3}}\raise26pt\hbox{$\ = \ 0$}
}}

While the original outline for finding tree-level Jacobi-satisfying numerators relied on manually inverting the propagator matrix and exploiting the residual gauge freedom to establish locality, it was not long before the community realized that the KLT matrix, or momentum kernel, does indeed represent an inversion of the propagator matrix relevant to finding Jacobi-satisfying numerators~\refs{\BCJfromKLT, \MomKer}.   The prescription is to define the masters as the half-ladder diagrams with external legs $k_1$, and $k_n$ as  fixed farthest rungs, allowing all permutations of legs $\{2,\ldots,n-1\}$, as in \figmast.    All such master numerators for all permutations without label  $n-1$ as the second to last argument are set to vanish, with the remaining $(n-3)!$ masters set to be 
\eqn\fivezero{
N_{1|\,\tau(23\ldots n-2) ,n-1\,|n} \equiv \sum_{\rho \in S_{n-3}} A(1,\rho(2,3\ldots,n-2),n,n-1) S[\rho(23\ldots n-2)\, | \,\tau(23\ldots n-2) ]_1\,.
}
All  numerators follow via Jacobi from these master numerators.   These numerators are manifestly non-local (although of course all physical observables have the appropriate poles).  All  poles belonging to the vanishing masters have been absorbed by the non-vanishing masters.  As pointed out in \BCJ, as long as the color-stripped amplitudes obey kinematic Jacobi relations on residues, one can find a generalized gauge transformation (cf.\ ref.~\LeeUPY) consistent with Jacobi pushing these poles into the appropriate master numerators.  

One should expect that if such a local representation is always possible for a theory then there should be a closed form for local masters\foot{Locality of numerators for Jacobi-descendant graphs  when expressed in terms of local numerators from masters follows from the color-stripped amplitudes satisfying Jacobi on all poles~\BCJ.}.  Indeed, the authors of \DuTBC\ present  such  a closed-form construction for the NLSM, making the  key-insight that  the symmetric $(n-2)! \times (n-2)!$ form of the momentum kernel has the necessary freedom to allow for locality, while recognizing the need for an off-shell regulation.  The naive on-shell attempt fails as BCJ relations imply that
\eqn\fiveone{
 \sum_{\rho \in S_{n-2}} S[ \sigma(23\ldots n-1) \, | \, \rho(23 \ldots n-1) ]_1 A(1,\rho(2,3, \ldots, n-1), n) =0\,.
}
Indeed, this on-shell failure was realized in the first symmetric $(n-2)! \times (n-2)!$
construction of a momentum kernel~\BohrMomKer.  The authors of
ref.~\BohrMomKer\
proposed a regulation of such a $S[\cdot | \cdot ]_1$ in the practice of building
gravitational amplitudes symmetrically from a KK basis.  They did so by
regulating the product of the symmetric sum with $1/s_{12 \ldots n-1}$ to cancel
an overall $s_{12 \ldots n-1}$ and then taking the
appropriate $s_{12 \ldots n-1}\to0$ limit. The authors of \DuTBC\ invoke such
a regulation in building local NLSM master numerators proposing
\eqnn\mastnum
$$\eqalignno{
N_{1|\sigma(23\ldots n-1)|n} &\equiv \lim_{s_{12\ldots n-1}\rightarrow 0} s_{12\ldots n-1}^{-1} \sum_{\rho \in S_{n-2}} S[\sigma(23\ldots n-1) | \rho(23\ldots n-1)]_1 \cr
& \ \ \ \ \ \ \ \ \ \ \ \  \ \ \ \ \ \ \ \ \ \ \ \ \ \ \ \ \times 
A_{\rm NLSM}(1,\rho(2,3,\ldots,n-1),n) \ . &\mastnum
}$$
Locality does indeed arise when the scattering amplitudes are expressed in an appropriate 
basis of Mandelstam variables as we now describe.
In a similar fashion as in the Berends--Giele \BerendsME\ description
of NLSM amplitudes \BGTrnka,
one can extend NLSM amplitudes to an
off-shell momentum $k_n^2 \neq 0$ by
using an overcomplete set of Mandelstam variables $s_{ij}$ with $1\leq
i<j\leq n{-}1$. Accordingly, the sum over $\rho$ in \mastnum\ gives rise to
an overall factor of $s_{12\ldots n-1}= \sum_{i<j}^{n-1}s_{ij}$ which cancels the
propagator $s_{12\ldots n-1}^{-1}$. This, in turn, yields a well-defined expression
upon the elimination $s_{1,n-1} \rightarrow -\sum_{i<j}^{n-2}s_{ij} - \sum_{i=2}^{n-2} s_{i,n-1}$
which implements the on-shell limit $s_{12\ldots n-1}\rightarrow 0$.

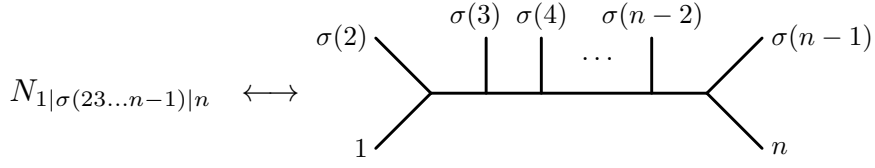
\ifig\figmast{Master diagrams with respect to Jacobi relations which are associated
with the master numerators $N_{1|\sigma(23\ldots n-1)|n}$ defined by \mastnum.}
{\hbox{
\raise18pt\hbox{$N_{1|\sigma(23\ldots n-1)|n} \ \ \longleftrightarrow \enskip$}%
{\epsfxsize=0.45\hsize\epsfbox{figmast.1}}
}}

The four- and six-point numerators \mastnum\ corresponding to the
amplitudes \NLSMfour\ and \NLSMsix\ read
\eqnn\locnumfour
\eqnn\locnumsix
$$\eqalignno{
N_{1|23|4} &= \pi^2 s_{1 2} (s_{1 3} + s_{2 3}) 
&\locnumfour \cr
N_{1|2345|6}  &= -\pi^4 s_{12} (s_{13}+s_{23}) (s_{14}+s_{24}+s_{34}) (s_{15}+s_{25}+s_{35}+s_{45})   \ .&\locnumsix
   }$$
The numerators of any descendant cubic diagram are then simply defined by a sequence of 
kinematic Jacobi identities as well as antisymmetry under flips of cubic vertices. Remarkably, the four- and six-point numerators in \locnumfour\ and \locnumsix\ coincide with the diagonal entries of the KLT matrix \mkerA. On these grounds, we propose the following all-multiplicity formula for NLSM master numerators \mastnum,
\eqn\conj{
N_{1|\rho(23\ldots 2k-1)|2k} = (-1)^{k} S[ \rho(23\ldots 2k{-}1) | \rho(23\ldots 2k{-}1) ]_1 \ .
}
We have verified their validity through multiplicity $2k=8$. While indeed surprising, one should note that ref.~\DuTBC\ arrived at a master numerator representation involving sums over permutations of  KLT matrix elements.

The form of the local numerators depends on the choice of implementing momentum conservation in
$A_{\rm NLSM}(\ldots)$. For instance, the on-shell equivalent expression $-\pi^2
s_{13}$ for $A_{\rm NLSM}(1,2,3,4)$ instead of $\pi^2(s_{12}+s_{23})$ yields $-
\pi^2 s_{1 2} s_{1 3}$ for the numerator $N_{1|23|4} $ instead of $\pi^2 s_{1 2}
(s_{1 3} + s_{2 3}) $. The six-point numerator \locnumsix\ is obtained from the NLSM amplitude \NLSMsix\ after converting the Mandelstam invariants into the nine-element basis $\{s_{1 2}, s_{13}, s_{1 4}, s_{2 3}, s_{2 4}, s_{2 5}, s_{3 4}, s_{3 5}, s_{45}\}$. The $n$-point generalization of this basis choice applicable to \conj\ reads $\{s_{ij} \ | \ 1\leq i<j \leq n{-}1 \ \& \ (i,j) \neq (1,n{-}1)\}$.

At generic multiplicity, the connection between color-ordered NLSM
amplitudes and master numerators \mastnum\ is captured by doubly-partial 
amplitudes in \defDP\ and \ftlim,
\eqn\nlsmmaster{
A_{\rm NLSM}(\Sigma(1,2,\ldots,n-1),n) =
\!\!\!\sum_{\rho \in S_{n-2}}\!\!\! m[\Sigma(12\ldots n-1)n | 1\rho(23\ldots n-1) n]
N_{1|\rho(23\ldots n-1)|n}  \ ,
}
which leads to the following expressions at four points,
\eqnn\fourex
$$\eqalignno{
A_{\rm NLSM}(1,2,3,4) &= A_{\rm NLSM}(3,2,1,4) = \frac{ N_{1|23|4} }{s_{12}} + \frac{ N_{1|23|4} - N_{1|32|4}  }{s_{23} }
 \cr
A_{\rm NLSM}(1,3,2,4) &= A_{\rm NLSM}(2,3,1,4) =
 \frac{ N_{1|32|4} }{s_{13}} + \frac{ N_{1|32|4} - N_{1|23|4}  }{s_{23} }
 &\fourex \cr
A_{\rm NLSM}(2,1,3,4) &= A_{\rm NLSM}(3,1,2,4) = -  \frac{ N_{1|23|4} }{s_{12}}  -  \frac{ N_{1|32|4} }{s_{13}} \ .
}$$
One can verify from \locnumfour\ that the
four-point amplitude \NLSMfour\ is correctly reproduced. 

As pointed out in \symmBCJ\ one can always symmetrize
Jacobi-satisfying numerators to arrive at a crossing symmetric
function for the generically dressed half-ladder topology in a
manner that preserves linear relations (like Jacobi).  One can note
that fully crossing-symmetric local numerators of \DuTBC\ were
arrived at by evaluating the Berends--Giele currents in the pion
parameterization scheme. This exemplifies how field redefinitions in
the context of Lagrangians amount to diagram by diagram
reparametrizations of scattering amplitudes which are often referred
to as generalized gauge transformations.  While the amplitudes
themselves are invariant under field redefinition and gauge choice
-- the local weights of individual graphs are not. This suggests an
interesting connection between generalized gauge transformations at
the amplitude level and field redefinitions as well as gauge choice
in the context of Lagrangians.

The above prescription to convert amplitudes subject to BCJ relations into local
and Jacobi-satisfying kinematic numerators is straightforwardly applied to
the full-fledged abelian disk integrals \abdisk. As in \mastnum, we
define $(n-2)!$ master numerators associated with the half-ladder diagrams in \figmast
\eqnn\mastZ
$$\eqalignno{
&N^{\times}_{1|\sigma(23\ldots n-1)|n}(\alpha') \equiv (\ap)^{2-n}\!\!\! \lim_{s_{12\ldots n-1}\rightarrow 0} s_{12\ldots n-1}^{-1}  &\mastZ \cr
& \ \ \ \ \ \times \sum_{\rho \in S_{n-2}} S[\sigma(23\ldots n-1) | \rho(23\ldots n-1)]_1 
 Z_{\times}(1,\rho(2,3,\ldots,n-1),n)\,, 
}$$
and corresponding $\ap$-corrected NLSM amplitudes:
\eqn\xnlsmmaster{
(\ap)^{2-n}Z_{\times}(\Sigma(1,\ldots,n-1),n) =
\!\!\!\sum_{\rho \in S_{n-2}}\!\!\! m[\Sigma(1\ldots n-1)n | 1\rho(2\ldots n-1) n]
N^\times_{1|\rho(23\ldots n-1)|n}(\ap).
}
For example, the $\ap$-corrections to the abelian four-point disk
integrals \fourab\ generalize the master numerator \locnumfour\ to (recalling that $\sigma_2 \equiv {\ap^2 \over 2} (s_{12}^2 + s_{13}^2 + s_{23}^2)$
and $\sigma_3 \equiv - \ap^3 s_{12}s_{23} s_{13}$)
\eqn\fourxalpha{
N^{\times}_{1|23|4} (\alpha')= \pi^2 s_{1 2} (s_{1 3} + s_{2 3}) \times \Big(1 +
\frac{1}{2} \zeta_2 \sigma_2 + \zeta_3 \sigma_3 + \frac{3}{10} \zeta_2^2\sigma_2^2  + {\cal O}(\alpha'^5) \Big) \ .
}
The analogous six-point corrections to \locnumsix\ at the order of $\zeta_6$ and
$\zeta_4 \zeta_3$ are attached as ancillary files to the arXiv submission of this
work; they yield \sigsixtwo\ and \sigsixthree\ according to \xnlsmmaster.

Of course, the construction of $\alpha'$-dependent numerators can be truncated to
each desired order in $\alpha'$ and refined to any MZV which is conjecturally
linearly independent over $\Bbb Q$. For example, the $\alpha'^2 \zeta_2$ and
$\alpha'^3 \zeta_3$ orders of $N^{\times}_{1|\sigma(23\ldots n-1)|n}(\alpha')$
generate BCJ numerators in an effective theory where the NLSM interactions are
supplemented by higher-derivative corrections with an extra $\alpha'^2 \zeta_2
\partial^4$ and $\alpha'^3 \zeta_3 \partial^6$, respectively.

\newsec Conclusions and outlook

In this paper, we interpret the disk integrals in open-string tree-level amplitudes as the
S-matrix of a collection of putative scalar effective theories we refer to as $Z$-theory. Our key result \preview\ establishes that 
the low-energy 
limit of abelian $Z$-theory amplitudes yields the $n$-point amplitudes of the NLSM at order $\ap^{n-2}$, 
while the next orders define new higher-derivative corrections which admit color-dual representations. Using this setup,
we obtain $\ap$-corrections to the local BCJ-satisfying numerators recently
identified by Du and Fu \DuTBC, and indeed, a novel all-multiplicity expression \entireexpression\ for the master numerators of the NLSM.

Given the high orders of $\ap$ involved in extracting the NLSM amplitudes and corrections
from the $Z$-amplitudes, the straightforward organization of the string-theory calculations presented in
this paper is not optimized to probe high multiplicities. Here, we chose to instead emphasize the relationship between
open-string predictions, abelian $Z$-amplitudes, and explicit $\ap$-corrections.
 In future work \refs{\BGap, \SAbZ}, efficient calculations will be
addressed by the corollaries of monodromy relations presented in \MaUM\
and a Berends--Giele recursion for the $\ap$-expansion of
non-abelian disk integrals using an extension of the method described in~\FTlimit.

Towards identifying patterns within $Z$-theory, we find ourselves encouraged to 
investigate the relevant higher-derivative corrections to the Lagrangian
description of the NLSM which reproduces the higher
$\ap$-corrections of the amplitudes discussed in this work.  
Preliminary considerations suggest that these are not the only higher-derivative
corrections consistent with color-kinematics. Accordingly, additional guiding principles 
may need to be invoked to arrive at the selection rules and the patterns of MZVs realized
by worldsheets of disk topology.  It has not escaped our 
notice that apprehending such guiding principles could indeed prove a fruitful line of inquiry. 

In addition to providing higher derivative corrections to the NLSM we have through
double copy, {\it en passant}, generated predictions
for a set of higher-derivative corrections to Born--Infeld, and its supersymmetric partners
including Volkov--Akulov from the fermionic sector~ \refs{\HeVFI,\CachazoNJL}. It would be interesting to 
contrast with higher-derivative corrected Born--Infeld-type theories existing in the literature, 
cf.\ the set of self-dual theories constructed in ref.~\ChemissanyYV. 

\bigskip \noindent{\bf Acknowledgements:} We are grateful to Nima Arkani-Hamed, Zvi
Bern, James Drummond, Song He, Yu-tin Huang, Henrik Johansson, Renata Kallosh, and
Radu Roiban for combinations of inspiring discussions and related collaboration.  We would
like to especially thank Radu Roiban for valuable comments on an initial draft.
The authors would like to thank IAS at Princeton where this work was initiated
as well as the Higgs Center, Nordita and the Isaac Newton Institute, and relevant
program organizers, for providing stimulating atmosphere, support, and hospitality
through the ``QCD meets Gravity'', ``Aspects of Amplitudes'', and ``Gravity,
Twistors, and Amplitudes'' programs, respectively.  JJMC is supported by the
European Research Council under ERC-STG-639729, {\it Strategic Predictions for Quantum
Field Theories\/}. CRM is supported by a University Research Fellowship from the
Royal Society, and gratefully acknowledges support from NSF grant number PHY
1314311 and the Paul Dirac Fund and EPSRC grant no EP/K032208/1.

\appendix{A}{Factorization plausibility}

\subsec{Via double copy}
\par\subseclab\newsubsecA

\noindent
Here we sketch an argument why one might expect all participants of a double-copy scattering amplitude to support factorization on its massless poles based on quite general double-copy properties.  Up to the orders in $\ap$ explored through eight points this is obviously the case between multiplicities for the color-ordered abelian $Z$-amplitudes presented here.  The interested reader can verify these properties continue to hold for higher point amplitudes and higher derivative $\ap$ corrections, including non-trivial Chan--Paton factors, using the doubly-ordered $Z$-amplitudes presented in \BGap.

Let us use $X = Y  \otimes Z$ to represent the massless-adjoint BCJ double-copy construction 
of scattering amplitudes where $Y$ and $Z$ may encode states but such encoding is absolutely independent.  The massless states of $X$ are taken to be the outer product of the states of $Y$ and $Z$.  We take as given that $X$ factorizes on massless channels.  This means that the residues of $X$ on any of its massless channels are given as the sum over $X$ states of products of lower-point $X$ amplitudes evaluated with on-shell momenta.  We take that
$Y$ represents the kinematic information from color-stripped amplitudes which also factorizes on massless channels.  This means that $Y \otimes \phi^3$ represents color-dressed scattering amplitudes in a $Y$ theory. 
For the case of abelian $Z$-functions presented here, $X$ would represent the abelianized open string, $Y$ would represent supersymmetric Yang--Mills, and $Z$ would represent the $\ap$ dependent kinematic information associated with abelian $Z$-functions.   
 
Since $X$ factorizes, and $Y$ factorizes, we can expect the numerator weights of $Z$ to factorize on massless channels up to only terms that vanish due to algebraic properties satisfied by $Y$.  Since we expect $Y$ and $Z$ states to be independent, we expect any such putative zeroes to represent generalized gauge freedom -- terms that vanish simply by virtue of a double copy with any theory that satisfies Jacobi and anti-symmetry.  As such color-dressed abelian $Z$-amplitudes: $Z\otimes \phi^3$ should factorize since the biadjoint scalar amplitudes obey Jacobi and anti-symmetry in each color-weight.   Color-stripping these color-dressed $Z$-amplitudes to color-ordered $Z$-amplitudes should in turn be independent of any such zeros, as the zeros can not care about what gauge group was being stripped.  It should follow that any such color-ordered $Z$-amplitudes will factorize.

This argument neglects the interesting, but irrelevant to the order-by-order finite $\ap$ exploration of the putative massless effective field-theory, consideration of the factorization on  the infinite tower of higher-spin massive modes, which is entirely contained within the $Z$-factor of the open superstring.

\subsec {Via worldsheet correlation considerations}
\par\subseclab\newsubsecB

\noindent
In this section we argue that the higher-mass dimension extensions of the NLSM
encoded in the $\ap$-expansion of $Z$-theory amplitudes preserve unitarity. The
desired factorization on the massless poles follow from a worldsheet model that contains
$Z$-theory along with extra terms which will be argued to decouple in our organization of the effective-field-theory regime.
This worldsheet model for $Z$-theory amplitudes is based on open-string vertex operators\foot{A string-theory 
realization of these vertex operators with Kac--Moody currents of $SU(2)$ has been given in \GreenGA\
through a toroidal compactifications along with worldsheet boundary condensates.}
\eqn\vertexJ{
V_a(z) = J_a(z) e^{ik\cdot x(z)}
}
for the Goldstone bosons associated with a trivial Chan--Paton generator. 
The indices of the Kac--Moody currents $J_a(z)$ refer to the adjoint representation
of some Lie-algebra with traceless generators $t_a$. Such currents are well-known from the heterotic
string \GrossRR\ and their tree-level correlation functions have been studied in \frenkel
\eqnn\hetcorr
$$\eqalignno{
&\ \ \ \langle V_{a_1}(z_1) V_{a_2}(z_2) \ldots V_{a_n}(z_n) \rangle = \prod_{i<j}^n |z_{ij}|^{\ap
s_{ij}} \bigg\{ \sum_{\rho \in S_{n-1} } { {\rm Tr}(t_{a_1}t_{a_{\rho(2)}} \ldots t_{a_{\rho(n)}}) \over z_{1\rho(2)} z_{\rho(23)} \ldots z_{\rho(n-1,n)} z_{\rho(n),1} }  &\hetcorr \cr
& \ \ \ + \sum_{j=2}^n c_j \sum_{\rho \in S_{n-1} } { {\rm Tr}(t_{a_1}t_{a_{\rho(2)}} \ldots t_{a_{\rho(j)}}) {\rm Tr}(t_{a_{\rho(j+1)}} \ldots t_{a_{\rho(n)}}) \over (z_{1\rho(2)} z_{\rho(23)} \ldots z_{\rho(j-1,j)} z_{\rho(j),1} ) \
(z_{\rho(j+1,j+2)} \ldots z_{\rho(n-1,n)} z_{\rho(n,j+1)} )}  + {\cal O}({\rm Tr}^3) \bigg\} 
}$$
for some rational numbers $c_j$ irrelevant to the subsequent arguments. As indicated by 
$+ {\cal O}({\rm Tr}^3)$ in the second line, the full correlation function involves products of up to $\lfloor n/2 \rfloor$
traces each of which is accompanied by a cycle of $z_{ij}^{-1}$.

By the first line of \hetcorr, the color-dressed version of the $Z$-theory amplitude \abdisk\ can be reproduced from the
vertex operators \vertexJ,
\eqnn\modtrace
$$\eqalignno{
&\sum_{\rho \in S_{n-1}}  {\rm Tr}(t_{a_1}t_{a_{\rho(2)}} \ldots t_{a_{\rho(n)}}) 
Z_{\times}(1,\rho(2,3,\ldots,n))  + {\cal O}({\rm Tr}^2)  &\modtrace
 \cr
 &= \ap^{n-3} \sum_{\rho \in S_{n-1}}  \int_{D(1\rho(23\ldots n))} {\dz_1 \ \dz_2
\ \cdots \ \dz_n \over {\rm vol}(SL(2,\Bbb R))}\,
 \langle V_{a_1}(z_1) V_{a_2}(z_2) \ldots V_{a_n}(z_n) \rangle   \ .
}$$
However, the correlator \hetcorr\ yields multitrace corrections to \modtrace\ where the integrands are built from
multiple cycles of $z_{ij}^{-1}$ instead of the single-cycle in \Zintdef. Such multicycle integrands spoil uniform
transcendentality of $Z$-amplitudes discussed in section \sectwofive\ as for instance seen in the four-point example
\eqn\multiexample{
\ap  \sum_{\rho \in S_3}
\int \limits_{D(1\rho(234))} {\dz_1 \ \dz_2\ \dz_3  \ \dz_4 \over {\rm vol}(SL(2,\Bbb R))}
 { \prod_{i<j}^4 |z_{ij}|^{\ap s_{ij}}  \over z_{12} z_{21}   z_{34} z_{43}} = {\ap s_{23} \over 1-\ap s_{12}} Z_{\times}(1,2,3,4) \ ,
}
also see section 4 of \OSEym\ for the general product of two cycles. By the geometric-series 
expansion ${1\over 1-\ap s_{ij}}= \sum_{n=0}^{\infty} (\ap s_{ij})^n$ and uniform transcendentality
of $ Z_{\times}(1,2,3,4)$, the coefficients of $\ap^n$ in \multiexample\ only contain MZVs of weight $w<n$. The same property 
extends to the entire multitrace corrections in \modtrace: For each term in the $\ap$-expansion of \modtrace, the multitrace terms
can be distinguished from the $Z$-amplitudes in the single-trace sector by the weight-drop in their accompanying MZVs. 

In the effective-field-theory viewpoint driving the present manuscript, factorization properties of \modtrace\ 
implied by the worldsheet model have to hold order by 
order in $\ap$ and separately along with any MZV which is conjecturally linearly independent over $\Bbb Q$. Restricting the unitarity analysis to the terms of uniform transcendentality, we conclude that $Z$-amplitudes factorize correctly on their massless poles.

\appendix{B}{The six-point $\alpha'^3 \zeta_3$ correction to the NLSM 
}

\noindent
In this appendix, we display the subleading $\alpha'$-correction to the NLSM model
as obtained from the abelian six-point disk integral at the order $\ap^7$:
\eqnn\sigsixthree
$$\eqalignno{
 Z_{\times}(1,2,3,4,5,6) \, \big|_{\alpha'^7}&=
 \pi^4 \zeta_3 \Big\{
-\frac{s_{12} s_{23} (s_{12} + s_{23})^2 (s_{45} + s_{56})}{s_{123}}
-s_{12}^2 s_{23} s_{34} - 3 s_{12} s_{23}^2 s_{34} \cr
& \! \! \! \! \! \! \! \! \! \! \! \! \! \! \! \! \! \! \! \! \! \! \! \! \! \! \! \! \! \!  - s_{12} s_{23} s_{34}^2 + s_{12}^3 s_{45} + 
 2 s_{12}^2 s_{23} s_{45} + 2 s_{12} s_{23}^2 s_{56} - s_{12} s_{23} s_{34} s_{56} + 
 s_{12}^2 s_{34} s_{123}  \cr
& \! \! \! \! \! \! \! \! \! \! \! \! \! \! \! \! \! \! \! \! \! \! \! \! \! \! \! \! \! \!  + 2 s_{12} s_{23} s_{34} s_{123} + s_{12}^2 s_{23} s_{345} + s_{12} s_{34}^2 s_{123} + 
 3 s_{23} s_{34}^2 s_{123} + s_{34}^3 s_{123} + s_{12} s_{23} s_{45} s_{123} \cr
& \! \! \! \! \! \! \! \! \! \! \! \! \! \! \! \! \! \! \! \! \! \! \! \! \! \! \! \! \! \!   - s_{12} s_{23} s_{56} s_{234} + 
 3 s_{12}^2 s_{23} s_{234} + s_{12} s_{23}^2 s_{234} - s_{12}^2 s_{234}^2 + s_{12} s_{23} s_{234}^2 + 
 s_{12} s_{23} s_{56} s_{123} \cr
& \! \! \! \! \! \! \! \! \! \! \! \! \! \! \! \! \! \! \! \! \! \! \! \! \! \! \! \! \! \!  + s_{12}^2 s_{34} s_{234} + 2 s_{12} s_{23} s_{34} s_{234} + 
 s_{12} s_{34}^2 s_{234} + 4 s_{12} s_{23} s_{45} s_{234} + 4 s_{12} s_{34} s_{45} s_{234} \cr
& \! \! \! \! \! \! \! \! \! \! \! \! \! \! \! \! \! \! \! \! \! \! \! \! \! \! \! \! \! \!   - 
 s_{12}^2 s_{123} s_{234} - 2 s_{12} s_{23} s_{123} s_{234} - s_{12} s_{23} s_{45} s_{345} - 
 s_{12}^2 s_{123} s_{345} - 2 s_{12} s_{23} s_{123} s_{345} \cr
& \! \! \! \! \! \! \! \! \! \! \! \! \! \! \! \! \! \! \! \! \! \! \! \! \! \! \! \! \! \!  - s_{12} s_{123}^2 s_{234} + s_{12} s_{123} s_{234}^2 - 
 s_{12} s_{34} s_{123} s_{345} + s_{12} s_{34} s_{56} s_{345} - 2 s_{12} s_{34} s_{123} s_{234} \cr
& \! \! \! \! \! \! \! \! \! \! \! \! \! \! \! \! \! \! \! \! \! \! \! \! \! \! \! \! \! \!  - 
 s_{12} s_{34} s_{234} s_{345} - s_{12} s_{234} s_{345}^2 - 4 s_{12} s_{23} s_{234} s_{345} + s_{12} s_{23} s_{345}^2 -
  s_{12}^2 s_{234} s_{345} \cr
& \! \! \! \! \! \! \! \! \! \! \! \! \! \! \! \! \! \! \! \! \! \! \! \! \! \! \! \! \! \!  + s_{12} s_{234}^2 s_{345} - s_{12} s_{234}^3 - 
 s_{12}^2 s_{34} s_{56}- 2 s_{12} s_{23} s_{45} s_{56} - 
 2 s_{12} s_{45} s_{234}^2 - \frac{1}{2} s_{12} s_{45} s_{123}^2 \cr
& \! \! \! \! \! \! \! \! \! \! \! \! \! \! \! \! \! \! \! \! \! \! \! \! \! \! \! \! \! \!   - \frac{1}{2} s_{12} s_{45} s_{345}^2 + 
 s_{123}^2 s_{234} s_{345} + \frac{1}{2} s_{123}^3 s_{234} + \frac{1}{2} s_{123}^3 s_{345} + {\rm cyc}(1,2,\ldots,6) \Big\} \ .
&\sigsixthree
}$$

\bigskip
\bigskip

\ninerm
\listrefs
 
\bye

%% file: harvmacM.tex


\input amssym.tex 

\def\unredoffs{}
\tolerance=1000\hfuzz=2pt
\catcode`\@=11 
\ifx\hyperdef\UNd@FiNeD\def\hyperdef#1#2#3#4{#4}\def\hyperref#1#2#3#4{#4}\def\href#1#2{#2}\fi
\magnification=1200\unredoffs\baselineskip=16pt plus 2pt minus 1pt
%

%
%

{\count255=\time\divide\count255 by 60 \xdef\hourmin{\number\count255}
 \multiply\count255 by-60\advance\count255 by\time
 \xdef\hourmin{\hourmin:\ifnum\count255<10 0\fi\the\count255}
}
\def\date{\number\day.\number\month.\number\year\ at \hourmin}


\def\nolabels{\def\wrlabeL##1{}\def\eqlabeL##1{}\def\reflabeL##1{}}
\def\writelabels{\def\wrlabeL##1{\leavevmode\vadjust{\rlap{\smash%
{\line{{\escapechar=` \hfill\rlap{\sevenrm\hskip.03in\string##1}}}}}}}%
\def\eqlabeL##1{{\escapechar-1\rlap{\sevenrm\hskip.05in\string##1}}}%
\def\reflabeL##1{\noexpand\llap{\noexpand\sevenrm\string\string\string##1}}}
\nolabels

\global\newcount\secno \global\secno=0
\global\newcount\meqno \global\meqno=1
\def\s@csym{}

\def\newsec#1\par{\global\advance\secno by1%
{\toks0{#1}\message{(\the\secno. \the\toks0)}}%
\global\subsecno=0\eqnres@t\let\s@csym\secsym\xdef\secn@m{\the\secno}\noindent
{\bf\hyperdef\hypernoname{section}{\the\secno}{\the\secno.} #1}%
\writetoca{{\string\hyperref{}{section}{\the\secno}{\bf \the\secno\quad}} {\bf #1}}\par%
\nobreak\medskip\nobreak\noindent\ignorespaces}
\def\eqnres@t{\xdef\secsym{\the\secno.}\global\meqno=1\bigbreak\bigskip}
\def\sequentialequations{\def\eqnres@t{\bigbreak}}\xdef\secsym{}

\global\newcount\subsecno \global\subsecno=0
\def\subsec#1\par{\global\advance\subsecno by1%
{\toks0{#1}\message{(\s@csym\the\subsecno. \the\toks0)}}%
\global\subsubsecno=0%
\ifnum\lastpenalty>9000\else\bigbreak\fi
\noindent{\it\hyperdef\hypernoname{subsection}{\secn@m.\the\subsecno}%
{\secn@m.\the\subsecno.} #1}\writetoca{\string\hskip1.45cm
{\string\hyperref{}{subsection}{\secn@m.\the\subsecno}{\secn@m.\the\subsecno.}}
{#1}}\par\nobreak\medskip\nobreak\noindent\ignorespaces}

\def\appendix#1#2{\global\meqno=1\global\subsecno=0\xdef\secsym{\hbox{#1.}}%
\bigbreak\bigskip\noindent{\bf Appendix \hyperdef\hypernoname{appendix}{#1}%
{#1.} #2}{\toks0{(#1. #2)}\message{\the\toks0}}%
\xdef\s@csym{#1.}\xdef\secn@m{#1}%
\writetoca{{\string\hyperref{}{appendix}{#1}{\bf {#1}\quad}} {\bf #2}}%
\par\nobreak\medskip\nobreak}

%
\def\checkm@de#1#2{\ifmmode{\def\f@rst##1{##1}\hyperdef\hypernoname{equation}%
{#1}{#2}}\else\hyperref{}{equation}{#1}{#2}\fi}
\def\eqnn#1{\DefWarn#1\xdef #1{(\noexpand\relax\noexpand\checkm@de%
{\s@csym\the\meqno}{\secsym\the\meqno})}%
\wrlabeL#1\writedef{#1\leftbracket#1}\global\advance\meqno by1}
\def\f@rst#1{\c@t#1a\em@ark}\def\c@t#1#2\em@ark{#1}
\def\eqna#1{\DefWarn#1\wrlabeL{#1$\{\}$}%
\xdef #1##1{(\noexpand\relax\noexpand\checkm@de%
{\s@csym\the\meqno\noexpand\f@rst{##1}1}{\hbox{$\secsym\the\meqno##1$}})}
\writedef{#1\numbersign1\leftbracket#1{\numbersign1}}\global\advance\meqno by1}
\def\eqn#1#2{\DefWarn#1%
\xdef #1{(\noexpand\hyperref{}{equation}{\s@csym\the\meqno}%
{\secsym\the\meqno})}$$#2\eqno(\hyperdef\hypernoname{equation}%
{\s@csym\the\meqno}{\secsym\the\meqno})\eqlabeL#1$$%
\writedef{#1\leftbracket#1}\global\advance\meqno by1}
\def\xeqn{\expandafter\xe@n}\def\xe@n(#1){#1}
\def\xeqna#1{\expandafter\xe@n#1}
\def\eqns#1{(\e@ns #1{\hbox{}})}
\def\e@ns#1{\ifx\UNd@FiNeD#1\message{eqnlabel \string#1 is undefined.}%
\xdef#1{(?.?)}\fi{\let\hyperref=\relax\xdef\next{#1}}%
\ifx\next\em@rk\def\next{}\else%
\ifx\next#1\xeqn#1\else\def\n@xt{#1}\ifx\n@xt\next#1\else\xeqna#1\fi
\fi\let\next=\e@ns\fi\next}

\def\DefWarn#1{\ifx\UNd@FiNeD#1\else
\immediate\write16{*** WARNING: the label \string#1 is already defined ***}\fi}
%
\newskip\footskip\footskip14pt plus 1pt minus 1pt 
\def\footnotefont{\ninepoint}\def\f@t#1{\footnotefont #1\@foot}
\def\f@@t{\baselineskip\footskip\bgroup\footnotefont\aftergroup\@foot\let\next}
\setbox\strutbox=\hbox{\vrule height9.5pt depth4.5pt width0pt}
\global\newcount\ftno \global\ftno=0
\def\foot{\global\advance\ftno by1\def\foot@rg{\hyperref{}{footnote}%
{\the\ftno}{\the\ftno}\xdef\foot@rg{\noexpand\hyperdef\noexpand\hypernoname%
{footnote}{\the\ftno}{\the\ftno}}}\footnote{$^{\foot@rg}$}}
%
%
%
\global\newcount\refno \global\refno=1
\newwrite\rfile
\def\ref{[\hyperref{}{reference}{\the\refno}{\the\refno}]\nref}
\def\nref#1{\DefWarn#1%
\xdef#1{[\noexpand\hyperref{}{reference}{\the\refno}{\the\refno}]}%
\writedef{#1\leftbracket#1}%
\ifnum\refno=1\immediate\openout\rfile=\jobname.refs\fi
\chardef\wfile=\rfile\immediate\write\rfile{\noexpand\item{[\noexpand\hyperdef%
\noexpand\hypernoname{reference}{\the\refno}{\the\refno}]\ }%
\reflabeL{#1\hskip.31in}\pctsign}\global\advance\refno by1\findarg}
\def\findarg#1#{\begingroup\obeylines\newlinechar=`\^^M\pass@rg}
{\obeylines\gdef\pass@rg#1{\writ@line\relax #1^^M\hbox{}^^M}%
\gdef\writ@line#1^^M{\expandafter\toks0\expandafter{\striprel@x #1}%
\edef\next{\the\toks0}\ifx\next\em@rk\let\next=\endgroup\else\ifx\next\empty%
\else\immediate\write\wfile{\the\toks0}\fi\let\next=\writ@line\fi\next\relax}}
\def\striprel@x#1{} \def\em@rk{\hbox{}}
\def\lref{\begingroup\obeylines\lr@f}
\def\lr@f#1#2{\DefWarn#1\gdef#1{\let#1=\UNd@FiNeD\ref#1{#2}}\endgroup\unskip}
\def\semi{;\hfil\break}
\def\addref#1{\immediate\write\rfile{\noexpand\item{}#1}} 
\def\listrefs{\vfill\supereject
\immediate\closeout\rfile\writestoppt
\baselineskip=\footskip\centerline{{\bf References}}\bigskip{\parindent=20pt%
\frenchspacing\escapechar=` \input \jobname.refs\vfill\eject}\nonfrenchspacing}
\def\startrefs#1{\immediate\openout\rfile=\jobname.refs\refno=#1}
\def\xref{\expandafter\xr@f}\def\xr@f[#1]{#1}
\def\refs#1{\count255=1[\r@fs #1{\hbox{}}]}
\def\r@fs#1{\ifx\UNd@FiNeD#1\message{reflabel \string#1 is undefined.}%
\nref#1{need to supply reference \string#1.}\fi%
\vphantom{\hphantom{#1}}{\let\hyperref=\relax\xdef\next{#1}}%
\ifx\next\em@rk\def\next{}%
\else\ifx\next#1\ifodd\count255\relax\xref#1\count255=0\fi%
\else#1\count255=1\fi\let\next=\r@fs\fi\next}
%

%
\newwrite\ffile\global\newcount\figno \global\figno=1
\def\fig{fig.~\hyperref{}{figure}{\the\figno}{\the\figno}\nfig}
\def\nfig#1{\DefWarn#1%
\xdef#1{fig.~\noexpand\hyperref{}{figure}{\the\figno}{\the\figno}}%
\writedef{#1\leftbracket fig.\noexpand~\xfig#1}%
\ifnum\figno=1\immediate\openout\ffile=\jobname.figs\fi\chardef\wfile=\ffile%
{\let\hyperref=\relax
\immediate\write\ffile{\noexpand\medskip\noexpand\item{Fig.\ %
\noexpand\hyperdef\noexpand\hypernoname{figure}{\the\figno}{\the\figno}. }
\reflabeL{#1\hskip.55in}\pctsign}}\global\advance\figno by1\findarg}
\def\xfig{\expandafter\xf@g}\def\xf@g fig.\penalty\@M\ {}
\def\figs#1{figs.~\f@gs #1{\hbox{}}}
\def\f@gs#1{{\let\hyperref=\relax\xdef\next{#1}}\ifx\next\em@rk\def\next{}\else
\ifx\next#1\xfig #1\else#1\fi\let\next=\f@gs\fi\next}
%
\def\figin{\epsfcheck\figin}\def\figins{\epsfcheck\figins}
\def\epsfcheck{\ifx\epsfbox\UnDeFiNeD
\message{(NO epsf.tex, FIGURES WILL BE IGNORED)}
\gdef\figin##1{\vskip2in}\gdef\figins##1{\hskip.5in}
\else\message{(FIGURES WILL BE INCLUDED)}%
\gdef\figin##1{##1}\gdef\figins##1{##1}\fi}
\def\DefWarn#1{}
\def\figinsert{\goodbreak\topinsert}
\def\ifig#1#2#3{\DefWarn#1\xdef#1{fig.~\the\figno}
\writedef{#1\leftbracket fig.\noexpand~\the\figno}%
\figinsert\figin{\centerline{#3}}
\smallskip
\leftskip=0pt \rightskip=0pt
\baselineskip12pt\noindent
{{\bf Fig.~\the\figno}\ \ninepoint #2}
\medskip
\global\advance\figno by1\par\endinsert}
\newwrite\lfile
{\escapechar-1\xdef\pctsign{\string\%}\xdef\leftbracket{\string\{}
\xdef\rightbracket{\string\}}\xdef\numbersign{\string\#}}
\def\writedefs{\immediate\openout\lfile=label.defs \def\writedef##1{%
{\let\hyperref=\relax\let\hyperdef=\relax\let\hypernoname=\relax
 \immediate\write\lfile{\string\def\string##1\rightbracket}}}}%
\def\writestop{\def\writestoppt{\immediate\write\lfile{\string\pageno
 \the\pageno\string\startrefs\leftbracket\the\refno\rightbracket
 \string\def\string\secsym\leftbracket\secsym\rightbracket
 \string\secno\the\secno\string\meqno\the\meqno}\immediate\closeout\lfile}}
\def\writestoppt{}\def\writedef#1{}

\def\seclab#1{\DefWarn#1%
\xdef #1{\noexpand\hyperref{}{section}{\the\secno}{\the\secno}}%
\writedef{#1\leftbracket#1}\wrlabeL{#1=#1}\par%
\nobreak\medskip\nobreak\noindent\ignorespaces}
\def\subseclab#1\par{\DefWarn#1%
\xdef #1{\noexpand\hyperref{}{subsection}{\the\secno.\the\subsecno}%
{\the\secno.\the\subsecno}}\writedef{#1\leftbracket#1}\wrlabeL{#1=#1}\par%
\nobreak\medskip\nobreak\noindent\ignorespaces}
\def\applab#1{\DefWarn#1%
\xdef #1{\noexpand\hyperref{}{appendix}{\secn@m}{\secn@m}}%
\writedef{#1\leftbracket#1}\wrlabeL{#1=#1}}
\newwrite\tfile \def\writetoca#1{}
\def\leaderfill{\leaders\hbox to 1em{\hss.\hss}\hfill}
\def\writetoc{\immediate\openout\tfile=\jobname.toc
   \def\writetoca##1{{\edef\next{\write\tfile{\noindent ##1
   \string\leaderfill{
   \string\hyperref{}{page}{\noexpand\number\pageno}%
   {\noexpand\number\pageno}} \par}}\next}}
}
\newread\ch@ckfile
\def\listtoc{\immediate\closeout\tfile\immediate\openin\ch@ckfile=\jobname.toc
\ifeof\ch@ckfile\message{no file \jobname.toc, no table of contents this pass}%
\else\closein\ch@ckfile\centerline{\bf Contents}\nobreak\medskip%
{\baselineskip=16pt\footnotefont\parskip=0pt\catcode`\@=11\input\jobname.toc
\catcode`\@=12\bigbreak\bigskip}\fi}
\catcode`\@=12 
\font\ninerm=cmr9 \font\sixrm=cmr6 \font\ninei=cmmi9 \font\sixi=cmmi6
\font\ninesy=cmsy9 \font\sixsy=cmsy6 \font\ninebf=cmbx9
\font\nineit=cmti9 \font\ninesl=cmsl9 \skewchar\ninei='177
\skewchar\sixi='177 \skewchar\ninesy='60 \skewchar\sixsy='60
\def\ninepoint{\def\rm{\fam0\ninerm}
\textfont0=\ninerm \scriptfont0=\sixrm \scriptscriptfont0=\fiverm
\textfont1=\ninei \scriptfont1=\sixi \scriptscriptfont1=\fivei
\textfont2=\ninesy \scriptfont2=\sixsy \scriptscriptfont2=\fivesy
\textfont\itfam=\ninei \def\it{\fam\itfam\nineit}\def\sl{\fam\slfam\ninesl}%
\textfont\bffam=\ninebf \def\bf{\fam\bffam\ninebf}\rm}
%
\hyphenation{anom-aly anom-alies coun-ter-term coun-ter-terms}

\global\newcount\subsubsecno \global\subsubsecno=0
\def\subsubsec#1\par{\global\advance\subsubsecno by1%
{\toks0{#1}\message{(\the\secno\the\subsecno\the\subsubsecno. \the\toks0)}}%
\ifnum\lastpenalty>9000\else\bigbreak\fi
\noindent{\it\hyperdef\hypernoname{subsubsection}{\the\secno.\the\subsecno\the\subsubsecno}%
{\the\secno.\the\subsecno.\the\subsubsecno.} #1}
\par\nobreak\medskip\nobreak\noindent\ignorespaces}

\def\DefWarn#1{}
\def\tikzcaption#1#2{\DefWarn#1\xdef#1{Fig.~\the\figno}
\writedef{#1\leftbracket Fig.\noexpand~\the\figno}%
{
\smallskip
\leftskip=20pt \rightskip=20pt \baselineskip12pt\noindent
{{\bf Fig.~\the\figno}\ \ninepoint #2}
\bigskip
\global\advance\figno by1 \par}}

\def\ntoalpha#1{%
\ifcase#1%
@%
\or A\or B\or C\or D\or E\or F\or G\or H\or I
\fi
}

\global\newcount\appno \global\appno=1
\def\applab#1{\xdef #1{\ntoalpha\appno}\writedef{#1\leftbracket#1}\wrlabeL{#1=#1}
\global\advance\appno by1}

\def\preprint#1 #2\par{\rightline{\vbox{\baselineskip12pt\hbox{#1}\hbox{#2}}}\vskip2cm}
%
\def\title#1\par{\centerline{\bf #1}\nopagenumbers\pageno=0}
\def\author#1\par{\bigskip\bigskip\centerline{#1}}

\newcount\addressno

\def\email#1#2{
\footnote{\null}{\kern-\parindent \llap{$^#1$\hskip1pt}email: #2}}

\def\startcenter{%
  \par
  \begingroup
  \leftskip=0pt plus 1fil
  \rightskip=\leftskip
  \parindent=0pt
  \parfillskip=0pt
}
\def\stopcenter{\endgroup}

\def\address{\bigskip%
  \ifnum\the\addressno=0\else\stopcenter\endgroup\fi
  \advance\addressno by 1%
  \begingroup
  \startcenter
  \it
  \obeylines
  \addressAux
}
\def\addressAux#1{#1}

\def\abstract{\stopcenter\endgroup\bigskip\bigskip\noindent}

\def\Dsl{\,\raise.15ex\hbox{/}\mkern-13.5mu D} 
\def\dsl{\raise.15ex\hbox{/}\kern-.57em\partial}
 
\def\boxeqn#1{\vcenter{\vbox{\hrule\hbox{\vrule\kern3pt\vbox{\kern3pt
	\hbox{${\displaystyle #1}$}\kern3pt}\kern3pt\vrule}\hrule}}}


\def\ap{{\alpha^{\prime}}}

\def\s{{\sigma}}

\def\half{{1\over 2}}

\def\({\left(}
\def\){\right)}
\def\dz{{\rm d}z}



\def\len#1{{%
\def\Dlen{\left|\mkern-1mu #1\mkern -0.5mu\right|}%
\def\Sslen{\left|\mkern-1.3mu #1\mkern -1.3mu\right|}%
\def\SSlen{\left|\mkern-2.8mu #1\mkern-1.3mu\right|}%
\mathchoice{\Dlen}{\Dlen}{\Sslen}{\SSlen}}}

\def\sfrac#1/#2{\kern.1em\raise.5ex\hbox{\the\scriptfont0 #1}%
\kern-.1em/\kern-.15em\lower.25ex\hbox{\the\scriptfont0 #2}}

\font\tenshuffle=shuffle10 \font\sevenshuffle=shuffle7 \font\fiveshuffle=shuffle7 at 5pt
\def\shuffle{{%
\def\Dshuffle{\mathbin{\hbox{\tenshuffle\char'001}}}%
\def\Sshuffle{\mathbin{\hbox{\sevenshuffle\char'001}}}%
\def\SSshuffle{\mathbin{\hbox{\fiveshuffle\char'001}}}%
\mathchoice{\Dshuffle}{\Dshuffle}{\Sshuffle}{\SSshuffle}}}


\def\qed{\hbox{\hskip 3pt
\vbox{\hrule\hbox to 7pt{\vrule height 7pt\hfill\vrule}
\hrule}}\hskip3pt}

\overfullrule=0pt\relax

\frenchspacing

\newread\instream \openin\instream= label.defs
\ifeof\instream \message{No labels in advance yet. Wait till next pass.}
\else \closein\instream \input label.defs
\fi
\writedefs

\def\arXiv:#1].{\hepthStrip#1 \nil}
\def\hepthStrip#1 #2\nil{\href{http://arxiv.org/abs/#1}{arXiv:#1 #2\unskip}].}

%% file: refs.tex
\lref\BernBX{
  Z.~Bern, A.~De Freitas and H.~L.~Wong,
  ``On the coupling of gravitons to matter,''
Phys.\ Rev.\ Lett.\  {\bf 84}, 3531 (2000).
[hep-th/9912033].
}

\lref\white{
  C.D.~White,
  ``Exact solutions for the biadjoint scalar field,''
[arXiv:1606.04724 [hep-th]].
}

\lref\GreenGA{
  M.~B.~Green and M.~Gutperle,
  ``Symmetry breaking at enhanced symmetry points,''
Nucl.\ Phys.\ B {\bf 460}, 77 (1996).
[hep-th/9509171].
}

\lref\deRooXV{
  M.~de Roo and M.~G.~C.~Eenink,
  ``The Effective action for the four point functions in Abelian open superstring theory,''
JHEP {\bf 0308}, 036 (2003).
[hep-th/0307211].
}

\lref\LeeUPY{
  S.~Lee, C.~R.~Mafra and O.~Schlotterer,
  ``Non-linear gauge transformations in $D=10$ SYM theory and the BCJ duality,''
JHEP {\bf 1603}, 090 (2016).
[arXiv:1510.08843 [hep-th]].
}

\lref\BIref{
  R.~R.~Metsaev, M.~Rakhmanov and A.~A.~Tseytlin,
  ``The {Born-Infeld} Action as the Effective Action in the Open Superstring Theory,''
Phys.\ Lett.\ B {\bf 193}, 207 (1987).
}

\lref\CachazoHCA{
  F.~Cachazo, S.~He and E.~Y.~Yuan,
  ``Scattering of Massless Particles in Arbitrary Dimensions,''
Phys.\ Rev.\ Lett.\  {\bf 113}, no. 17, 171601 (2014).
[arXiv:1307.2199 [hep-th]].
}

\lref\BerendsME{
	F.A.~Berends and W.T.~Giele,
  	``Recursive Calculations for Processes with n Gluons,''
	Nucl.\ Phys.\ B {\bf 306}, 759 (1988).
}

\lref\frenkel{
  I.~Frenkel and M.~Zhu, ``Vertex operator algebras associated to representations of affine and Virasoro algebras,'' 
  Duke Math J.\ 66, 123 (1992). \semi 
L.~Dolan and P.~Goddard,
  ``Current Algebra on the Torus,''
Commun.\ Math.\ Phys.\  {\bf 285}, 219 (2009).
[arXiv:0710.3743 [hep-th]].
   }
   
\lref\GrossRR{
  D.~J.~Gross, J.~A.~Harvey, E.~J.~Martinec and R.~Rohm,
  ``Heterotic String Theory. 2. The Interacting Heterotic String,''
Nucl.\ Phys.\ B {\bf 267}, 75 (1986)..
}   

\lref\BGTrnka{
  K.~Kampf, J.~Novotny and J.~Trnka,
  ``Recursion relations for tree-level amplitudes in the $SU(N)$ nonlinear sigma model,''
Phys.\ Rev.\ D {\bf 87}, no. 8, 081701 (2013).
[arXiv:1212.5224 [hep-th]].
\semi
K.~Kampf, J.~Novotny and J.~Trnka,
  ``Tree-level Amplitudes in the Nonlinear Sigma Model,''
JHEP {\bf 1305}, 032 (2013).
[arXiv:1304.3048 [hep-th]].
}
\lref\PSBCJ{
	C.R.~Mafra, O.~Schlotterer and S.~Stieberger,
	``Explicit BCJ Numerators from Pure Spinors,''
	JHEP {\bf 1107}, 092 (2011).
	[arXiv:1104.5224 [hep-th]].
	\semi
 C.~R.~Mafra and O.~Schlotterer,
  ``Berends-Giele recursions and the BCJ duality in superspace and components,''
JHEP {\bf 1603}, 097 (2016).
[arXiv:1510.08846 [hep-th]].
}
\lref\BGSym{
	F.A.~Berends and W.T.~Giele,
	``Multiple Soft Gluon Radiation in Parton Processes,''
	Nucl.\ Phys.\ B {\bf 313}, 595 (1989).
}
\lref\DPellis{
	F.~Cachazo, S.~He and E.Y.~Yuan,
	``Scattering of Massless Particles: Scalars, Gluons and Gravitons,''
	JHEP {\bf 1407}, 033 (2014).
	[arXiv:1309.0885 [hep-th]].
}
\lref\CHY{
	F.~Cachazo, S.~He and E.Y.~Yuan,
 	``Scattering equations and Kawai-Lewellen-Tye orthogonality,''
	Phys.\ Rev.\ D {\bf 90}, no. 6, 065001 (2014).
	[arXiv:1306.6575 [hep-th]].
}

\lref\CasaliVTA{
  E.~Casali, Y.~Geyer, L.~Mason, R.~Monteiro and K.~A.~Roehrig,
  ``New Ambitwistor String Theories,''
JHEP {\bf 1511}, 038 (2015).
[arXiv:1506.08771 [hep-th]].
}

\lref\dolan{
	L.~Dolan and P.~Goddard,
	``Proof of the Formula of Cachazo, He and Yuan for Yang-Mills Tree Amplitudes in Arbitrary Dimension,''
	JHEP {\bf 1405}, 010 (2014).
	[arXiv:1311.5200 [hep-th]].
}
\lref\KKsym{
	R.~Kleiss and H.~Kuijf,
	``Multi - Gluon Cross-sections and Five Jet Production at Hadron Colliders,''
	Nucl.\ Phys.\ B {\bf 312}, 616 (1989).
}

\lref\BI{
  M.~Born and L.~Infeld,
  ``Foundations of the new field theory,''
Proc.\ Roy.\ Soc.\ Lond.\ A {\bf 144}, 425 (1934).
}
\lref\Schrodinger{
  E.~Schr\"odinger,
  ``Contributions to Born's New Theory of the Electromagnetic Field,''
Proc.\ Roy.\ Soc.\ Lond.\ A {\bf 150}, 465 (1935).
\semi
  P.A.M.~Dirac,
  ``An Extensible model of the electron,''
Proc.\ Roy.\ Soc.\ Lond.\ A {\bf 268}, 57 (1962).
\semi
  D.~V.~Volkov and V.~P.~Akulov,
  ``Is the Neutrino a Goldstone Particle?,''
Phys.\ Lett.\ B {\bf 46}, 109 (1973).
\semi
  S.~Deser and C.~Teitelboim,
  ``Duality Transformations of Abelian and Nonabelian Gauge Fields,''
Phys.\ Rev.\ D {\bf 13}, 1592 (1976).
\semi
  M.~K.~Gaillard and B.~Zumino,
  ``Duality Rotations for Interacting Fields,''
Nucl.\ Phys.\ B {\bf 193}, 221 (1981).
\semi
  S.~Cecotti and S.~Ferrara,
  ``Supersymmetric Born-infeld Lagrangians,''
Phys.\ Lett.\ B {\bf 187}, 335 (1987).
\semi
  G.~W.~Gibbons and D.~A.~Rasheed,
  ``Electric - magnetic duality rotations in nonlinear electrodynamics,''
Nucl.\ Phys.\ B {\bf 454}, 185 (1995).
[hep-th/9506035].
\semi
  J.~Bagger and A.~Galperin,
  ``A New Goldstone multiplet for partially broken supersymmetry,''
Phys.\ Rev.\ D {\bf 55}, 1091 (1997).
[hep-th/9608177].
\semi
  M.~K.~Gaillard and B.~Zumino,
  ``Nonlinear electromagnetic selfduality and Legendre transformations,''
In *Cambridge 1997, Duality and supersymmetric theories* 33-48.
[hep-th/9712103].
\semi
  S.~V.~Ketov,
  ``A Manifestly N=2 supersymmetric Born-Infeld action,''
Mod.\ Phys.\ Lett.\ A {\bf 14}, 501 (1999).
[hep-th/9809121].
\semi
  S.~V.~Ketov,
  ``Many faces of Born-Infeld theory,''
[hep-th/0108189].
\semi
  M.~Rocek and A.~A.~Tseytlin,
  ``Partial breaking of global D = 4 supersymmetry, constrained superfields, and three-brane actions,''
Phys.\ Rev.\ D {\bf 59}, 106001 (1999).
[hep-th/9811232].
\semi
  A.~A.~Tseytlin,
  ``Born-Infeld action, supersymmetry and string theory,''
In *Shifman, M.A. (ed.): The many faces of the superworld* 417-452.
[hep-th/9908105].
\semi
  S.~M.~Kuzenko and S.~Theisen,
  ``Supersymmetric duality rotations,''
JHEP {\bf 0003}, 034 (2000).
[hep-th/0001068].
\semi
  S.~M.~Kuzenko and S.~Theisen,
  ``Nonlinear selfduality and supersymmetry,''
Fortsch.\ Phys.\  {\bf 49}, 273 (2001).
[hep-th/0007231].
\semi
  S.~Bellucci, E.~Ivanov and S.~Krivonos,
  ``Towards the complete N=2 superfield Born-Infeld action with partially broken N=4 supersymmetry,''
Phys.\ Rev.\ D {\bf 64}, 025014 (2001).
[hep-th/0101195].
\semi
  S.~Bellucci, E.~Ivanov and S.~Krivonos,
  ``N=2 and N=4 supersymmetric Born-Infeld theories from nonlinear realizations,''
Phys.\ Lett.\ B {\bf 502}, 279 (2001).
[hep-th/0012236].
\semi
  E.~A.~Ivanov and B.~M.~Zupnik,
  ``New representation for Lagrangians of selfdual nonlinear electrodynamics,''
[hep-th/0202203].
\semi
  E.~A.~Ivanov and B.~M.~Zupnik,
  ``New approach to nonlinear electrodynamics: Dualities as symmetries of interaction,''
Phys.\ Atom.\ Nucl.\  {\bf 67}, 2188 (2004), [Yad.\ Fiz.\  {\bf 67}, 2212 (2004)].
[hep-th/0303192].
}

\lref\HenneauxGG{
  M.~Henneaux and C.~Teitelboim,
  ``Dynamics of Chiral (Selfdual) $P$ Forms,''
Phys.\ Lett.\ B {\bf 206}, 650 (1988).
}

\lref\KKLance{
	V.~Del Duca, L.J.~Dixon and F.~Maltoni,
	``New color decompositions for gauge amplitudes at tree and loop level,''
	Nucl.\ Phys.\ B {\bf 571}, 51 (2000).
	[hep-ph/9910563].
}
\lref\kkbg{
	C.H.~Fu, Y.J.~Du and B.~Feng,
  	``An algebraic approach to BCJ numerators,''
	JHEP {\bf 1303}, 050 (2013).
	[arXiv:1212.6168 [hep-th]].
}

\lref\BlumleinCF{
  J.~Blumlein, D.~J.~Broadhurst and J.~A.~M.~Vermaseren,
  ``The Multiple Zeta Value Data Mine,''
Comput.\ Phys.\ Commun.\  {\bf 181}, 582 (2010).
[arXiv:0907.2557 [math-ph]].
}

\lref\BjerrumBohrXE{
  N.~E.~J.~Bjerrum-Bohr, P.~H.~Damgaard, H.~Johansson and T.~Sondergaard,
  ``Monodromy--like Relations for Finite Loop Amplitudes,''
JHEP {\bf 1105}, 039 (2011).
[arXiv:1103.6190 [hep-th]].
}

\lref\BjerrumBohrRD{
  N.~E.~J.~Bjerrum-Bohr, P.~H.~Damgaard and P.~Vanhove,
  ``Minimal Basis for Gauge Theory Amplitudes,''
Phys.\ Rev.\ Lett.\  {\bf 103}, 161602 (2009).
[arXiv:0907.1425 [hep-th]].
}
\lref\loopMonodromy{
  P.~Tourkine and P.~Vanhove,
  ``Higher-loop amplitude monodromy relations in string and gauge theory,''
[arXiv:1608.01665 [hep-th]].
}

\lref\StiebergerHQ{
  S.~Stieberger,
  ``Open \& Closed vs. Pure Open String Disk Amplitudes,''
[arXiv:0907.2211 [hep-th]].
}

\lref\MomKer{
	N.~E.~J.~Bjerrum-Bohr, P.~H.~Damgaard, T.~Sondergaard and P.~Vanhove,
	``The Momentum Kernel of Gauge and Gravity Theories,''
	JHEP {\bf 1101}, 001 (2011).
	[arXiv:1010.3933 [hep-th]].
}
\lref\Ztheory{
	J.~Broedel, O.~Schlotterer and S.~Stieberger,
	``Polylogarithms, Multiple Zeta Values and Superstring Amplitudes,''
	Fortsch.\ Phys.\  {\bf 61}, 812 (2013).
	[arXiv:1304.7267 [hep-th]].
}

\lref\BroedelAZA{
  J.~Broedel, O.~Schlotterer, S.~Stieberger and T.~Terasoma,
  ``All order $\alpha^{\prime}$-expansion of superstring trees from the Drinfeld associator,''
Phys.\ Rev.\ D {\bf 89}, no. 6, 066014 (2014).
[arXiv:1304.7304 [hep-th]].
}

\lref\nptTreeI{
	C.R.~Mafra, O.~Schlotterer and S.~Stieberger,
	``Complete N-Point Superstring Disk Amplitude I. Pure Spinor Computation,''
	Nucl.\ Phys.\ B {\bf 873}, 419 (2013).
	[arXiv:1106.2645 [hep-th]].
}
\lref\nptTreeII{
	C.R.~Mafra, O.~Schlotterer and S.~Stieberger,
	``Complete N-Point Superstring Disk Amplitude II. Amplitude
	and Hypergeometric Function Structure,''
	Nucl.\ Phys.\ B {\bf 873}, 461 (2013).
	[arXiv:1106.2646 [hep-th]].
}

\lref\StiebergerRR{
  S.~Stieberger,
  ``Constraints on Tree-Level Higher Order Gravitational Couplings in Superstring Theory,''
Phys.\ Rev.\ Lett.\  {\bf 106}, 111601 (2011).
[arXiv:0910.0180 [hep-th]].
}

\lref\MasonSVA{
  L.~Mason and D.~Skinner,
  ``Ambitwistor strings and the scattering equations,''
JHEP {\bf 1407}, 048 (2014).
[arXiv:1311.2564 [hep-th]].
}
\lref\AdamoTSA{
  T.~Adamo, E.~Casali and D.~Skinner,
  ``Ambitwistor strings and the scattering equations at one loop,''
JHEP {\bf 1404}, 104 (2014).
[arXiv:1312.3828 [hep-th]].
}
\lref\GeyerBJA{
  Y.~Geyer, L.~Mason, R.~Monteiro and P.~Tourkine,
  ``Loop Integrands for Scattering Amplitudes from the Riemann Sphere,''
Phys.\ Rev.\ Lett.\  {\bf 115}, no. 12, 121603 (2015).
[arXiv:1507.00321 [hep-th]].
}
\lref\GeyerJCH{
  Y.~Geyer, L.~Mason, R.~Monteiro and P.~Tourkine,
  ``One-loop amplitudes on the Riemann sphere,''
JHEP {\bf 1603}, 114 (2016).
[arXiv:1511.06315 [hep-th]].
}
\lref\CardonaBPI{
  C.~Cardona and H.~Gomez,
  ``Elliptic scattering equations,''
JHEP {\bf 1606}, 094 (2016).
[arXiv:1605.01446 [hep-th]].
}
\lref\CardonaWCR{
  C.~Cardona and H.~Gomez,
  ``CHY-Graphs on a Torus,''
[arXiv:1607.01871 [hep-th]].
}
\lref\chyTwoLoop{
  Y.~Geyer, L.~Mason, R.~Monteiro and P.~Tourkine,
  ``Two-Loop Scattering Amplitudes from the Riemann Sphere,''
[arXiv:1607.08887 [hep-th]].
}

\lref\nptMethod{
	C.R.~Mafra, O.~Schlotterer, S.~Stieberger and D.~Tsimpis,
	``A recursive method for SYM n-point tree amplitudes,''
	Phys.\ Rev.\ D {\bf 83}, 126012 (2011).
	[arXiv:1012.3981 [hep-th]].
}

\lref\KLT{
	H.~Kawai, D.C.~Lewellen and S.H.H.~Tye,
	``A Relation Between Tree Amplitudes of Closed and Open Strings,''
	Nucl.\ Phys.\ B {\bf 269}, 1 (1986).
}

\lref\BroedelRC{
  J.~Broedel and L.~J.~Dixon,
  ``Color-kinematics duality and double-copy construction for amplitudes from higher-dimension operators,''
JHEP {\bf 1210}, 091 (2012).
[arXiv:1208.0876 [hep-th]].
}

\lref\OprisaWU{
  D.~Oprisa and S.~Stieberger,
  ``Six gluon open superstring disk amplitude, multiple hypergeometric series and Euler-Zagier sums,''
[hep-th/0509042].
}

\lref\multigluon{
S.~Stieberger and T.~R.~Taylor,
  ``Multi-Gluon Scattering in Open Superstring Theory,''
Phys.\ Rev.\ D {\bf 74}, 126007 (2006).
[hep-th/0609175].
}

\lref\GreenFT{
  M.~B.~Green, J.~H.~Schwarz and L.~Brink,
  ``N=4 Yang-Mills and N=8 Supergravity as Limits of String Theories,''
Nucl.\ Phys.\ B {\bf 198}, 474 (1982).
}

\lref\mathMZV{
T.~Terasoma, ``Selberg integrals and multiple zeta values,'' Compositio Mathematica {\bf 133}, 1 (2002).
\semi
K.~Aomoto, ``Special values of hyperlogarithms and linear difference schemes,''
 Illinois J. Math. {\bf 34} (2), 191 (1990).
\semi
F.~Brown, ``Multiple zeta values and periods of moduli spaces ${\cal M}_{0,n}$,''
Ann. Sci. Ec. Norm. Super. {\bf 42} (4), 371 (2009),
[math/0606419].
}

\lref\psf{
 	N.~Berkovits,
	``Super-Poincare covariant quantization of the superstring,''
	JHEP {\bf 0004}, 018 (2000)
	[arXiv:hep-th/0001035].
}
\lref\Duhr{
	C.~Duhr, S.~Hoeche and F.~Maltoni,
	``Color-dressed recursive relations for multi-parton amplitudes,''
	JHEP {\bf 0608}, 062 (2006).
	[hep-ph/0607057].
}
\lref\BCJ{
	Z.~Bern, J.J.M.~Carrasco and H.~Johansson,
  	``New Relations for Gauge-Theory Amplitudes,''
	Phys.\ Rev.\ D {\bf 78}, 085011 (2008).
	[arXiv:0805.3993 [hep-ph]].
}	
\lref\BCJLoops{
  Z.~Bern, J.~J.~M.~Carrasco and H.~Johansson,
  ``Perturbative Quantum Gravity as a Double Copy of Gauge Theory,''
Phys.\ Rev.\ Lett.\  {\bf 105}, 061602 (2010).
[arXiv:1004.0476 [hep-th]].
}

\lref\BernYG{
  Z.~Bern, T.~Dennen, Y.~t.~Huang and M.~Kiermaier,
  ``Gravity as the Square of Gauge Theory,''
Phys.\ Rev.\ D {\bf 82}, 065003 (2010).
[arXiv:1004.0693 [hep-th]].
}

\lref\BernUF{
  Z.~Bern, J.~J.~M.~Carrasco, L.~J.~Dixon, H.~Johansson and R.~Roiban,
  ``Simplifying Multiloop Integrands and Ultraviolet Divergences of Gauge Theory and Gravity Amplitudes,''
Phys.\ Rev.\ D {\bf 85}, 105014 (2012).
[arXiv:1201.5366 [hep-th]].
\semi
Z.~Bern, S.~Davies, T.~Dennen and Y.~t.~Huang,
  ``Absence of Three-Loop Four-Point Divergences in N=4 Supergravity,''
Phys.\ Rev.\ Lett.\  {\bf 108}, 201301 (2012).
[arXiv:1202.3423 [hep-th]].
\semi
Z.~Bern, S.~Davies, T.~Dennen, A.~V.~Smirnov and V.~A.~Smirnov,
  ``Ultraviolet Properties of N=4 Supergravity at Four Loops,''
Phys.\ Rev.\ Lett.\  {\bf 111}, no. 23, 231302 (2013).
[arXiv:1309.2498 [hep-th]].
\semi
Z.~Bern, S.~Davies and T.~Dennen,
  ``Enhanced ultraviolet cancellations in ${\cal N}=5$ supergravity at four loops,''
Phys.\ Rev.\ D {\bf 90}, no. 10, 105011 (2014).
[arXiv:1409.3089 [hep-th]].
}

\lref\BerendsZP{
  F.~A.~Berends, W.~T.~Giele and H.~Kuijf,
  ``On relations between multi - gluon and multigraviton scattering,''
Phys.\ Lett.\ B {\bf 211}, 91 (1988).
}

\lref\BernKLT{
  Z.~Bern, L.~J.~Dixon, M.~Perelstein and J.~S.~Rozowsky,
  ``Multileg one loop gravity amplitudes from gauge theory,''
Nucl.\ Phys.\ B {\bf 546}, 423 (1999).
[hep-th/9811140].
}

\lref\gravUVTwo{
  Z.~Bern, L.~J.~Dixon, D.~C.~Dunbar, M.~Perelstein and J.~S.~Rozowsky,
  ``On the relationship between Yang-Mills theory and gravity and its implication for ultraviolet divergences,''
Nucl.\ Phys.\ B {\bf 530}, 401 (1998).
[hep-th/9802162].
\semi
  Z.~Bern, J.~J.~Carrasco, L.~J.~Dixon, H.~Johansson, D.~A.~Kosower and R.~Roiban,
  ``Three-Loop Superfiniteness of N=8 Supergravity,''
Phys.\ Rev.\ Lett.\  {\bf 98}, 161303 (2007).
[hep-th/0702112].
\semi
  Z.~Bern, J.~J.~M.~Carrasco, L.~J.~Dixon, H.~Johansson and R.~Roiban,
  ``Manifest Ultraviolet Behavior for the Three-Loop Four-Point Amplitude of N=8 Supergravity,''
Phys.\ Rev.\ D {\bf 78}, 105019 (2008).
[arXiv:0808.4112 [hep-th]].
\semi
  Z.~Bern, J.~J.~Carrasco, D.~Forde, H.~Ita and H.~Johansson,
  ``Unexpected Cancellations in Gravity Theories,''
Phys.\ Rev.\ D {\bf 77}, 025010 (2008).
[arXiv:0707.1035 [hep-th]].
\semi
  Z.~Bern, J.~J.~Carrasco, L.~J.~Dixon, H.~Johansson and R.~Roiban,
  ``The Ultraviolet Behavior of N=8 Supergravity at Four Loops,''
Phys.\ Rev.\ Lett.\  {\bf 103}, 081301 (2009).
[arXiv:0905.2326 [hep-th]].
}

\lref\LitseyJFA{
  S.~Litsey and J.~Stankowicz,
  ``Kinematic numerators and a double-copy formula for $N$=4 super-Yang-Mills residues,''
Phys.\ Rev.\ D {\bf 90}, no. 2, 025013 (2014).
[arXiv:1309.7681 [hep-th]].
\semi
  R.~Monteiro and D.~O'Connell,
  ``The Kinematic Algebras from the Scattering Equations,''
JHEP {\bf 1403}, 110 (2014).
[arXiv:1311.1151 [hep-th]].
\semi
  N.~E.~J.~Bjerrum-Bohr, J.~L.~Bourjaily, P.~H.~Damgaard and B.~Feng,
  ``Manifesting Color-Kinematics Duality in the Scattering Equation Formalism,''
[arXiv:1608.00006 [hep-th]].
}

\lref\OSEym{
  O.~Schlotterer,
  ``Amplitude relations in heterotic string theory and Einstein-Yang-Mills,''
[arXiv:1608.00130 [hep-th]].
}

\lref\ChyEym{
  F.~Cachazo, S.~He and E.~Y.~Yuan,
  ``Einstein-Yang-Mills Scattering Amplitudes From Scattering Equations,''
JHEP {\bf 1501}, 121 (2015).
[arXiv:1409.8256 [hep-th]].
}

\lref\FORM{
	J.A.M.~Vermaseren,
  	``New features of FORM,''
	[math-ph/0010025].
\semi
	J.~Kuipers, T.~Ueda, J.A.M.~Vermaseren and J.~Vollinga,
	``FORM version 4.0,''
	Comput.\ Phys.\ Commun.\  {\bf 184}, 1453 (2013).
	[arXiv:1203.6543 [cs.SC]].
}

\lref\FTlimit{
	C.R.~Mafra,
  	``Berends-Giele recursion for double-color-ordered amplitudes,''
	JHEP {\bf 1607}, 080 (2016).
	[arXiv:1603.09731 [hep-th]].
}

\lref\MafraKH{
	C.R.~Mafra and O.~Schlotterer,
  	``The Structure of n-Point One-Loop Open Superstring Amplitudes,''
	JHEP {\bf 1408}, 099 (2014).
	[arXiv:1203.6215 [hep-th]].
}

\lref\HuangTAG{
  Y.~t.~Huang, O.~Schlotterer and C.~Wen,
  ``Universality in string interactions,''
[arXiv:1602.01674 [hep-th]].
}

\lref\SchlottererNY{
  O.~Schlotterer and S.~Stieberger,
  ``Motivic Multiple Zeta Values and Superstring Amplitudes,''
J.\ Phys.\ A {\bf 46}, 475401 (2013).
[arXiv:1205.1516 [hep-th]].
}

\lref\DrummondVZ{
  J.~M.~Drummond and E.~Ragoucy,
  ``Superstring amplitudes and the associator,''
JHEP {\bf 1308}, 135 (2013).
[arXiv:1301.0794 [hep-th]].
}

\lref\WWW{ J. Broedel, O.~Schlotterer and S.~Stieberger,
{\tt http://mzv.mpp.mpg.de}
}

\lref\StiebergerHBA{
  S.~Stieberger and T.~R.~Taylor,
  ``Closed String Amplitudes as Single-Valued Open String Amplitudes,''
Nucl.\ Phys.\ B {\bf 881}, 269 (2014).
[arXiv:1401.1218 [hep-th]].
}

\lref\psweb{
  C.R.~Mafra, O.~Schlotterer,
http://www.damtp.cam.ac.uk/user/crm66/SYM/pss.html
}

\lref\DuTBC{
  Y.~J.~Du and C.~H.~Fu,
  ``Explicit BCJ numerators of nonlinear sigma model,''
[arXiv:1606.05846 [hep-th]].
}
\lref\scherk{
  J.~Scherk,
  ``Zero-slope limit of the dual resonance model,''
Nucl.\ Phys.\ B {\bf 31}, 222 (1971).
}
\lref\Frampton{
P. Frampton,
``Dual Resonance Models,'' Frontiers in Physics, Benjamin 1974.
}

\lref\GomezWZA{
  H.~Gomez and E.~Y.~Yuan,
  ``N-point tree-level scattering amplitude in the new Berkovits` string,''
JHEP {\bf 1404}, 046 (2014).
[arXiv:1312.5485 [hep-th]].
}

\lref\BerkovitsXBA{
  N.~Berkovits,
  ``Infinite Tension Limit of the Pure Spinor Superstring,''
JHEP {\bf 1403}, 017 (2014).
[arXiv:1311.4156 [hep-th]].
}

\lref\chendu{
  G.~Chen and Y.J.~Du,
  ``Amplitude Relations in Non-linear Sigma Model,''
JHEP {\bf 1401}, 061 (2014).
[arXiv:1311.1133 [hep-th]].
}

\lref\CachazoNLSM{
  F.~Cachazo, S.~He and E.~Y.~Yuan,
  ``Scattering Equations and Matrices: From Einstein To Yang-Mills, DBI and NLSM,''
JHEP {\bf 1507}, 149 (2015).
[arXiv:1412.3479 [hep-th]].
}

\lref\MafraGJA{
  C.R.~Mafra and O.~Schlotterer,
  ``Towards one-loop SYM amplitudes from the pure spinor BRST cohomology,''
Fortsch.\ Phys.\  {\bf 63}, no. 2, 105 (2015).
[arXiv:1410.0668 [hep-th]].
\semi
C.R.~Mafra and O.~Schlotterer,
  ``Two-loop five-point amplitudes of super Yang-Mills and supergravity in pure spinor superspace,''
JHEP {\bf 1510}, 124 (2015).
[arXiv:1505.02746 [hep-th]].
\semi
 S.~He, R.~Monteiro and O.~Schlotterer,
  ``String-inspired BCJ numerators for one-loop MHV amplitudes,''
JHEP {\bf 1601}, 171 (2016).
[arXiv:1507.06288 [hep-th]].
}

\lref\SelfDualBCJ{
	R.~Monteiro and D.~O'Connell,
  	``The Kinematic Algebra From the Self-Dual Sector,''
	JHEP {\bf 1107}, 007 (2011).
	[arXiv:1105.2565 [hep-th]].
}
\lref\ReutBook{
	C.~Reutenauer,
	``Free Lie Algebras'', London Mathematical Society Monographs, 1993.
}
\lref\fundBCJ{
	N.~E.~J.~Bjerrum-Bohr, P.~H.~Damgaard, B.~Feng and T.~Sondergaard,
  	``Proof of Gravity and Yang-Mills Amplitude Relations,''
	JHEP {\bf 1009}, 067 (2010).
	[arXiv:1007.3111 [hep-th]].
}
\lref\BohrMomKer{
	N.~E.~J.~Bjerrum-Bohr, P.~H.~Damgaard, B.~Feng and T.~Sondergaard,
	``Gravity and Yang-Mills Amplitude Relations,''
	Phys.\ Rev.\ D {\bf 82}, 107702 (2010).
	[arXiv:1005.4367 [hep-th]].
}

\lref\WittenNN{
  E.~Witten,
  ``Perturbative gauge theory as a string theory in twistor space,''
Commun.\ Math.\ Phys.\  {\bf 252}, 189 (2004).
[hep-th/0312171].
}
\lref\RoibanYF{
  R.~Roiban, M.~Spradlin and A.~Volovich,
  ``On the tree level S matrix of Yang-Mills theory,''
Phys.\ Rev.\ D {\bf 70}, 026009 (2004).
[hep-th/0403190].
}

\lref\BargheerGV{
  T.~Bargheer, S.~He and T.~McLoughlin,
  ``New Relations for 3-Dimensional Supersymmetric Scattering Amplitudes,''
Phys.\ Rev.\ Lett.\  {\bf 108}, 231601 (2012).
[arXiv:1203.0562 [hep-th]].
\semi
  J.~J.~M.~Carrasco, M.~Chiodaroli, M.~Günaydin and R.~Roiban,
  ``One-loop four-point amplitudes in pure and matter-coupled N $\leq$ 4 supergravity,''
JHEP {\bf 1303}, 056 (2013).
[arXiv:1212.1146 [hep-th]].
\semi
  Y.~t.~Huang and H.~Johansson,
  ``Equivalent D=3 Supergravity Amplitudes from Double Copies of Three-Algebra and Two-Algebra Gauge Theories,''
Phys.\ Rev.\ Lett.\  {\bf 110}, 171601 (2013).
[arXiv:1210.2255 [hep-th]].
\semi
  Y.~t.~Huang, H.~Johansson and S.~Lee,
  ``On Three-Algebra and Bi-Fundamental Matter Amplitudes and Integrability of Supergravity,''
JHEP {\bf 1311}, 050 (2013).
[arXiv:1307.2222 [hep-th]].
\semi
  M.~Chiodaroli, Q.~Jin and R.~Roiban,
  ``Color/kinematics duality for general abelian orbifolds of N=4 super Yang-Mills theory,''
JHEP {\bf 1401}, 152 (2014).
[arXiv:1311.3600 [hep-th]].
\semi
  H.~Johansson and A.~Ochirov,
  ``Pure Gravities via Color-Kinematics Duality for Fundamental Matter,''
JHEP {\bf 1511}, 046 (2015).
[arXiv:1407.4772 [hep-th]].
\semi
  M.~Chiodaroli, M.~Gunaydin, H.~Johansson and R.~Roiban,
  ``Scattering amplitudes in $ {\cal N}=2 $ Maxwell-Einstein and Yang-Mills/Einstein supergravity,''
JHEP {\bf 1501}, 081 (2015).
[arXiv:1408.0764 [hep-th]].
\semi
  H.~Johansson and A.~Ochirov,
  ``Color-Kinematics Duality for QCD Amplitudes,''
JHEP {\bf 1601}, 170 (2016).
[arXiv:1507.00332 [hep-ph]].
\semi
  M.~Chiodaroli, M.~Gunaydin, H.~Johansson and R.~Roiban,
  ``Spontaneously Broken Yang-Mills-Einstein Supergravities as Double Copies,''
[arXiv:1511.01740 [hep-th]].
\semi
  M.~Chiodaroli, M.~Gunaydin, H.~Johansson and R.~Roiban,
  ``Complete construction of magical, symmetric and homogeneous N=2 supergravities as double copies of gauge theories,''
Phys.\ Rev.\ Lett.\  {\bf 117}, no. 1, 011603 (2016).
[arXiv:1512.09130 [hep-th]].
\semi
  M.~Chiodaroli,
  ``Simplifying amplitudes in Maxwell-Einstein and Yang-Mills-Einstein supergravities,''
[arXiv:1607.04129 [hep-th]].
}

\lref\ChemissanyYV{
  W.~Chemissany, R.~Kallosh and T.~Ortin,
  ``Born-Infeld with Higher Derivatives,''
Phys.\ Rev.\ D {\bf 85}, 046002 (2012).
[arXiv:1112.0332 [hep-th]].
}

\lref\BCJfromKLT{
M. Kiermaier,  ``Gravity as the Square of Gauge Theory''\foot{{\tt   http://www.strings.ph.qmul.ac.uk/$\tilde{~}$theory/Amplitudes2010/Talks/MK2010.pdf }}, Amplitudes 2010, Queen Mary, University of London.
}

\lref\symmBCJ{
C.~H.~Fu, Y.~J.~Du and B.~Feng,
  ``Note on symmetric BCJ numerator,''
JHEP {\bf 1408}, 098 (2014).
[arXiv:1403.6262 [hep-th]].
\semi
  S.~G.~Naculich,
  ``Scattering equations and virtuous kinematic numerators and dual-trace functions,''
JHEP {\bf 1407}, 143 (2014).
[arXiv:1404.7141 [hep-th]].
}

\lref\MonteiroCDA{
  R.~Monteiro, D.~O'Connell and C.~D.~White,
  ``Black holes and the double copy,''
JHEP {\bf 1412}, 056 (2014).
[arXiv:1410.0239 [hep-th]].
\semi
  A.~Luna, R.~Monteiro, D.~O'Connell and C.~D.~White,
  ``The classical double copy for Taub-NUT spacetime,''
Phys.\ Lett.\ B {\bf 750}, 272 (2015).
[arXiv:1507.01869 [hep-th]].
\semi
  R.~Monteiro, D.~O'Connell and C.~D.~White,
  ``Gravity as a double copy of gauge theory: from amplitudes to black holes,''
Int.\ J.\ Mod.\ Phys.\ D {\bf 24}, no. 09, 1542008 (2015).
\semi
  A.~Luna, R.~Monteiro, I.~Nicholson, D.~O'Connell and C.~D.~White,
  ``The double copy: Bremsstrahlung and accelerating black holes,''
JHEP {\bf 1606}, 023 (2016).
[arXiv:1603.05737 [hep-th]].
\semi
  C.D.~White,
  ``Exact solutions for the biadjoint scalar field,''
[arXiv:1606.04724 [hep-th]].
}

\lref\BorstenBP{
  L.~Borsten, M.~J.~Duff, L.~J.~Hughes and S.~Nagy,
  ``Magic Square from Yang-Mills Squared,''
Phys.\ Rev.\ Lett.\  {\bf 112}, no. 13, 131601 (2014).
[arXiv:1301.4176 [hep-th]].
\semi
  A.~Anastasiou, L.~Borsten, M.~J.~Duff, L.~J.~Hughes and S.~Nagy,
  ``Super Yang-Mills, division algebras and triality,''
JHEP {\bf 1408}, 080 (2014).
[arXiv:1309.0546 [hep-th]].
\semi
  A.~Anastasiou, L.~Borsten, M.~J.~Duff, L.~J.~Hughes and S.~Nagy,
  ``A magic pyramid of supergravities,''
JHEP {\bf 1404}, 178 (2014).
[arXiv:1312.6523 [hep-th]].
\semi
  A.~Anastasiou, L.~Borsten, M.~J.~Duff, L.~J.~Hughes and S.~Nagy,
  ``Yang-Mills origin of gravitational symmetries,''
Phys.\ Rev.\ Lett.\  {\bf 113}, no. 23, 231606 (2014).
[arXiv:1408.4434 [hep-th]].
\semi
  A.~Anastasiou, L.~Borsten, M.~J.~Hughes and S.~Nagy,
  ``Global symmetries of Yang-Mills squared in various dimensions,''
JHEP {\bf 1601}, 148 (2016).
[arXiv:1502.05359 [hep-th]].
\semi
  L.~Borsten and M.~J.~Duff,
  ``Gravity as the square of Yang-Mills,''
Phys.\ Scripta {\bf 90}, 108012 (2015).
[arXiv:1602.08267 [hep-th]].
}

\lref\KampfVHA{
L.~Susskind and G.~Frye,
  ``Algebraic aspects of pionic duality diagrams,''
Phys.\ Rev.\ D {\bf 1}, 1682 (1970). \semi
H.~Osborn,
  ``Implications of adler zeros for multipion processes,''
Lett.\ Nuovo Cim.\  {\bf 2S1}, 717 (1969), [Lett.\ Nuovo Cim.\  {\bf 2}, 717 (1969)].
\semi
J.~R.~Ellis and B.~Renner,
  ``On the relationship between chiral and dual models,''
Nucl.\ Phys.\ B {\bf 21}, 205 (1970).
}

\lref\MaUM{
  Y.~J.~Du, Y.~X.~Chen and Q.~Ma,
  ``On Primary Relations at Tree-level in String Theory and Field Theory,''
JHEP {\bf 1202}, 061 (2012).
[arXiv:1109.0685 [hep-th]].
}

\lref\ChangZZA{
J.~A.~Cronin,
  ``Phenomenological model of strong and weak interactions in chiral U(3) x U(3),''
Phys.\ Rev.\  {\bf 161}, 1483 (1967).
\semi
S.~Weinberg,
  ``Dynamical approach to current algebra,''
Phys.\ Rev.\ Lett.\  {\bf 18}, 188 (1967).
\semi
S.~Weinberg,
  ``Nonlinear realizations of chiral symmetry,''
Phys.\ Rev.\  {\bf 166}, 1568 (1968).
\semi
L.~S.~Brown,
  ``Field Theory Of Chiral Symmetry,''
Phys.\ Rev.\  {\bf 163}, 1802 (1967).
\semi
  P.~Chang and F.~Gursey,
  ``Unified Formulation of Effective Nonlinear Pion-Nucleon Lagrangians,''
Phys.\ Rev.\  {\bf 164}, 1752 (1967)..
}

\lref\BGap{
  C.R.~Mafra and O.~Schlotterer,
  ``Non-abelian $Z$-theory: Berends-Giele recursion for the $\alpha'$-expansion of disk integrals,''
JHEP {\bf 1701}, 031 (2017).
[arXiv:1609.07078 [hep-th]].
}
\lref\gitrepo{
{\tt http://repo.or.cz/BGap.git}
}

\lref\SAbZ{
  J.~J.~M.~Carrasco, C.~R.~Mafra and O.~Schlotterer,
  ``Semi-abelian Z-theory: NLSM+$\phi^3$ from the open string,''
[arXiv:1612.06446 [hep-th]].
}

\lref\HeVFI{
  S.~He, Z.~Liu and J.~B.~Wu,
JHEP {\bf 1607}, 060 (2016).
[arXiv:1604.02834 [hep-th]].
}

\lref\CachazoNJL{
  F.~Cachazo, P.~Cha and S.~Mizera,
JHEP {\bf 1606}, 170 (2016).
[arXiv:1604.03893 [hep-th]].
}